\documentclass[useAMS,usenatbib]{mnras}

\usepackage{newtxtext,newtxmath}

\usepackage[T1]{fontenc}
\usepackage{ae,aecompl}
\usepackage[normalem]{ulem}
\usepackage{enumitem}
\usepackage{tabularx}
\usepackage{pdflscape}

\usepackage{graphicx}	
\usepackage{amsmath}	
\usepackage{hyperref}
\usepackage{academicons}
\usepackage[dvipsnames]{xcolor}
\usepackage{scalerel}
\usepackage{tikz}
\usetikzlibrary{svg.path}

\newcommand{\cc}{cm$^{-3}$}
\newcommand{\kmps}{km s$^{-1}$}

\newcommand{\hsfr}{SFR$10$}
\newcommand{\lsfr}{SFR$3$}
\newcommand{\sfr}{M$_{\odot}$ yr$^{-1}$}
\newcommand{\zsolar}{Z$_{\odot}$}
\newcommand{\mk}{$10^6$ K}

\definecolor{orcidlogocol}{HTML}{A6CE39}
\tikzset{
  orcidlogo/.pic={
    \fill[orcidlogocol] svg{M256,128c0,70.7-57.3,128-128,128C57.3,256,0,198.7,0,128C0,57.3,57.3,0,128,0C198.7,0,256,57.3,256,128z};
    \fill[white] svg{M86.3,186.2H70.9V79.1h15.4v48.4V186.2z}
                 svg{M108.9,79.1h41.6c39.6,0,57,28.3,57,53.6c0,27.5-21.5,53.6-56.8,53.6h-41.8V79.1z M124.3,172.4h24.5c34.9,0,42.9-26.5,42.9-39.7c0-21.5-13.7-39.7-43.7-39.7h-23.7V172.4z}
                 svg{M88.7,56.8c0,5.5-4.5,10.1-10.1,10.1c-5.6,0-10.1-4.6-10.1-10.1c0-5.6,4.5-10.1,10.1-10.1C84.2,46.7,88.7,51.3,88.7,56.8z};
  }
}

\newcommand\orcid[1]{\href{https://orcid.org/#1}{\mbox{\scalerel*{
\begin{tikzpicture}[yscale=-1,transform shape]
\pic{orcidlogo};
\end{tikzpicture}
}{|}}}}

\title[X-Ray Spectra of Hot CGM]{X-ray Spectra of Circumgalactic Medium Around Star-Forming Galaxies: Connecting Simulations to Observations}

\author[Vijayan \& Li]{
Aditi Vijayan\orcid{0000-0002-7714-2379}$^{1,2}$\thanks{E-mail: aditiv@rri.res.in},
Miao Li\orcid{0000-0003-0773-582X}$^{3,4}$ \thanks{miaoli@zju.edu.cn} 
\\
$^{1}$Raman Research Institute, Bangalore, 500080, India \\
$^{2}$Shanghai Astronomical Observatory, Chinese Academy of Sciences, 80 Nandan Road, Shanghai 200030, China\\
$^{3}$Department of Physics, Zhejiang University, 866 Yuhangtang Road, Hangzhou, 310058, China\\
$^{4}$Center for Computational Astrophysics, Flatiron Institute, 162 Fifth Avenue, New York, NY 10010, USA
}

\date{Accepted XXX. Received YYY; in original form ZZZ}

\pubyear{2021}

\begin{document}
\label{firstpage}
\pagerange{\pageref{firstpage}--\pageref{lastpage}}
\maketitle

\begin{abstract}

The hot component of the circum-galactic medium (CGM) around star forming galaxies is detected as diffuse X-ray emission. The X-ray spectra from the CGM depend on the temperature and metallicity of the emitting plasma, providing important information about the feeding and feedback of the galaxy. The observed spectra are commonly fitted using simple 1-Temperature (1-T) or 2-T models. However, the actual temperature distribution of the gas can be complex because of the interaction between galactic outflows and halo gas. Here we demonstrate this by analysing 3-D hydrodynamical simulations of the CGM with a realistic outflow model. We investigate the physical properties of the simulated hot CGM, which shows a broad distribution in density, temperature, and metallicity. By constructing and fitting the simulated spectra, we show that, while the 1-T and 2-T models are able to fit the synthesized spectra reasonably well, the inferred temperature(s) does not bear much physical meaning. Instead, we propose a log-normal distribution as a more physical model. The log-normal model better fits the simulated spectra while reproducing the gas temperature distribution. We also show that when the star formation rate is high, the spectra inside the bi-conical outflows are distinct from that outside, as outflows are generally hotter and more metal-enriched. Finally, we produce mock spectra for future missions with the eV-level spectral resolution, such as Athena, Lynx, HUBS and XRISM. 
\end{abstract}




\section{Introduction}

Circumgalactic medium (CGM), the multiphase gas surrounding a galaxy, has a symbiotic relationship with the galaxy. CGM serves as a reservoir providing fuel for star formation \citep[e.g.][] {Lehner+12, Lehner+13, Ford+14}, and the feedback from the galaxy leaves imprints on the CGM \citep{Oppenheimer&Dave08, Hummels+13,Fielding+17, Suresh+17,Li&Tonnesen19, Davies+20}. The CGM thus plays a critical role in galaxy evolution.

The X-ray emitting hot phase of the CGM, with a temperature $\gtrsim 10^6$ K, is also referred to as ``galactic corona''. 
X-ray observations have revealed a connection between the CGM and the galaxies hosting it. For star-forming galaxies, the X-ray luminosities of the hot CGM have positive scaling relations with the galaxy's mass and its star formation rate \citep{Strickland+04, Tullmann+06, Li+13, Wang+16}. Further, the X-ray emission is spatially correlated with star forming regions \citep{Grimes+05, Owen+09}. The extended X-ray emission has been detected up to several $\sim 10$s kpc in massive spirals \citep{Yamasaki+09, Anderson+11, Dai+12, Anderson+16, Bogdan+13, Bogdan+13B, Bogdan+17, Lopez+20}. Around early type galaxies, the hot halos have been detected out to $\sim 10s$ kpc and the X-ray luminosity is correlated with the optical luminosity \citep{Forman+85, OSullivan+01, Mulchaey+10}.

Theoretically, galaxy coronae can develop via two processes- cosmic accretion and galactic feedback. Gas infall into the dark matter (DM) halo from the inter-galactic medium (IGM) creates virial shocks, heating the gas to the virial temperature of the galaxy. If the halo is sufficiently massive (M$_{\rm vir} \gtrsim 10^{12} \rm{M}_\odot$), the shock-heated gas has a very long cooling time (e.g., $\sim 6$ Gyr for a $10^7$ K, $10^{-4}$ \cc gas and assuming cooling rate of $\sim 10^{-23}$ erg s$^{-1}$ cm$^{-3}$) and it forms a hot CGM that is in quasi-hydrostatic equilibrium with the gravitational potential of the DM halo \citep{McKee&Ostriker77, Silk77, Dekel+09,Keres+05}. Feedback from star formation and supermassive black holes releases copious of energy which can heat and accelerate the ambient inter-stellar medium (ISM) \citep{Fabian2010}. A fraction of the feedback energy and material can escape from the galaxy into the CGM, via galactic winds \citep{Sarkar+15, Li17a, Armillotta+17, Kim&Ostriker17, Vijayan+18, Li&Tonnesen19}. 

Galactic coronae are often studied through their X-ray emission spectra. The overall shape and the emission lines (if resolvable) are determined by the underlying gas properties, such as temperature, density, metallicity, etc. This information is then used to infer the galaxy formation process, such as gas feeding and feedback. It is thus critical to extract the gas properties from the spectra in a meaningful way.  
While interpreting the observed spectra, simple models for the temperature distribution of the underlying plasma are commonly assumed e.g., a 1-Temperature (1-T) or 2-T distribution. However, such assumptions for the CGM have rarely been been evaluated critically. In fact, there is reason to believe that the situation is more complex. Near the disk, where most of the emission arises, the hot galactic outflows undergo expansion, radiative cooling, interaction with pre-existing halo gas and other gas phases, etc. Thus, a 1-T or 2-T model may not be sufficient to capture the real temperature distribution of the gas.

In this paper, we evaluate the gas models used for interpreting the diffuse X-ray spectra from the CGM of a star forming galaxy. We use hydrodynamic simulations by \cite{Li&Tonnesen19} (referred to as ``L20'' hereafter), which model galactic outflows and their impact on the CGM for a Milky Way-mass galaxy. The simulations of L20 are suitable for X-ray studies because they incorporate a careful treatment of hot outflows taken from the results  from small-box simulations of supernovae-driven outflows from the ISM \citep{Li+17, li_bryan20}. We examine the X-ray emitting gas from the simulation and investigate the distribution of its temperature. We then construct synthesized X-ray spectra, and use them to evaluate the commonly used 1-T and 2-T models. We also make mock spectra for the upcoming X-ray satellites, Athena, Lynx, and HUBS. These telescopes will have unprecedented spectral resolution and sensitivity, which will transform our understanding about the CGM and thus galaxy formation. 

The paper is organised as follows. In Section \ref{sec:simulations} we discuss the general setup of simulations from L20 and in Section \ref{sec:results} we discuss the results from the simulations, including the outflow properties of the simulations (Section \ref{subsec:visualization_of_simulations}), the density and temperature distributions (Section \ref{subsec:rho-T-histos}), and the X-ray emission (Section \ref{subsec:X-ray_emission}). In Section \ref{subsec:1T-2T-spec} we produce synthesized spectra from the simulation data and compare it with 1-T and 2-T models. In Section \ref{sec:log-normal}, we propose a new model of temperature distribution and produce spectra from this model and compare with simulations. We then discuss spatial dependence of the emission spectra in Section \ref{sec:perp-spec}. We conclude by discussing this work with respect to previous works in Section \ref{sec:discussion}, and providing mock spectra for Chandra, Athena, Lynx and HUBS by convolving simulation spectra with instrument responses (Section \ref{subsec:predictions}). We summarize this paper in Section \ref{sec:summary}.

\begin{figure*}
	\includegraphics[width=\textwidth]{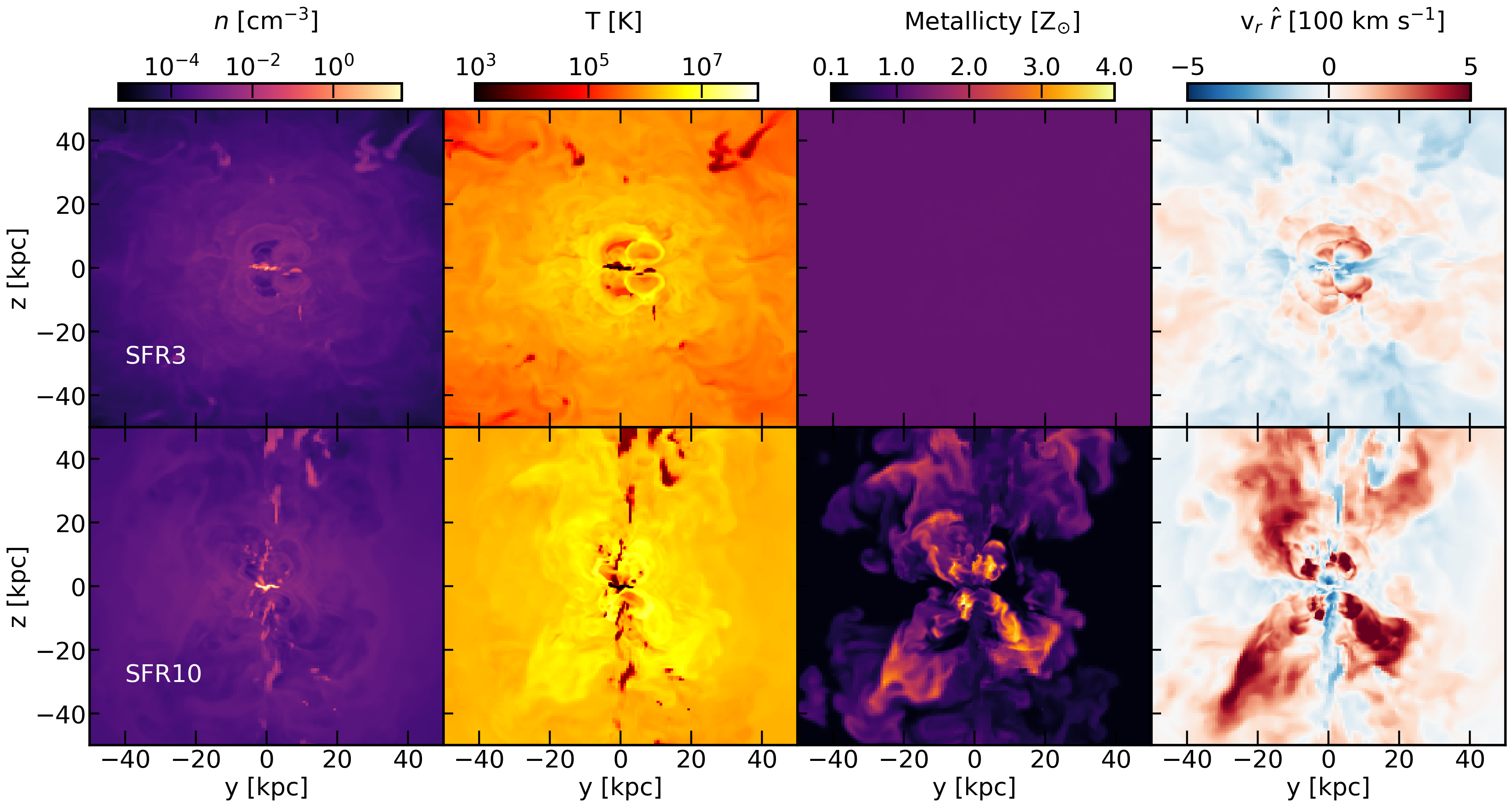}
    \caption{An edge view for $x=0$ slices from the simulations of \citet{Li&Tonnesen19} for \lsfr\ (top) and \hsfr\ (bottom) runs at $3$ and $1.3$ Gyr, respectively, showing density, temperature, metallicity, and radial velocity. \lsfr\ forms a large-scale fountain flows, while \hsfr\ has bi-conical outflows. Gas near the disk show a wide range of density and temperature, as a result of episodic outflows.}
\end{figure*}\label{fig:x-slices}

\begin{figure}
	\includegraphics[width=\columnwidth]{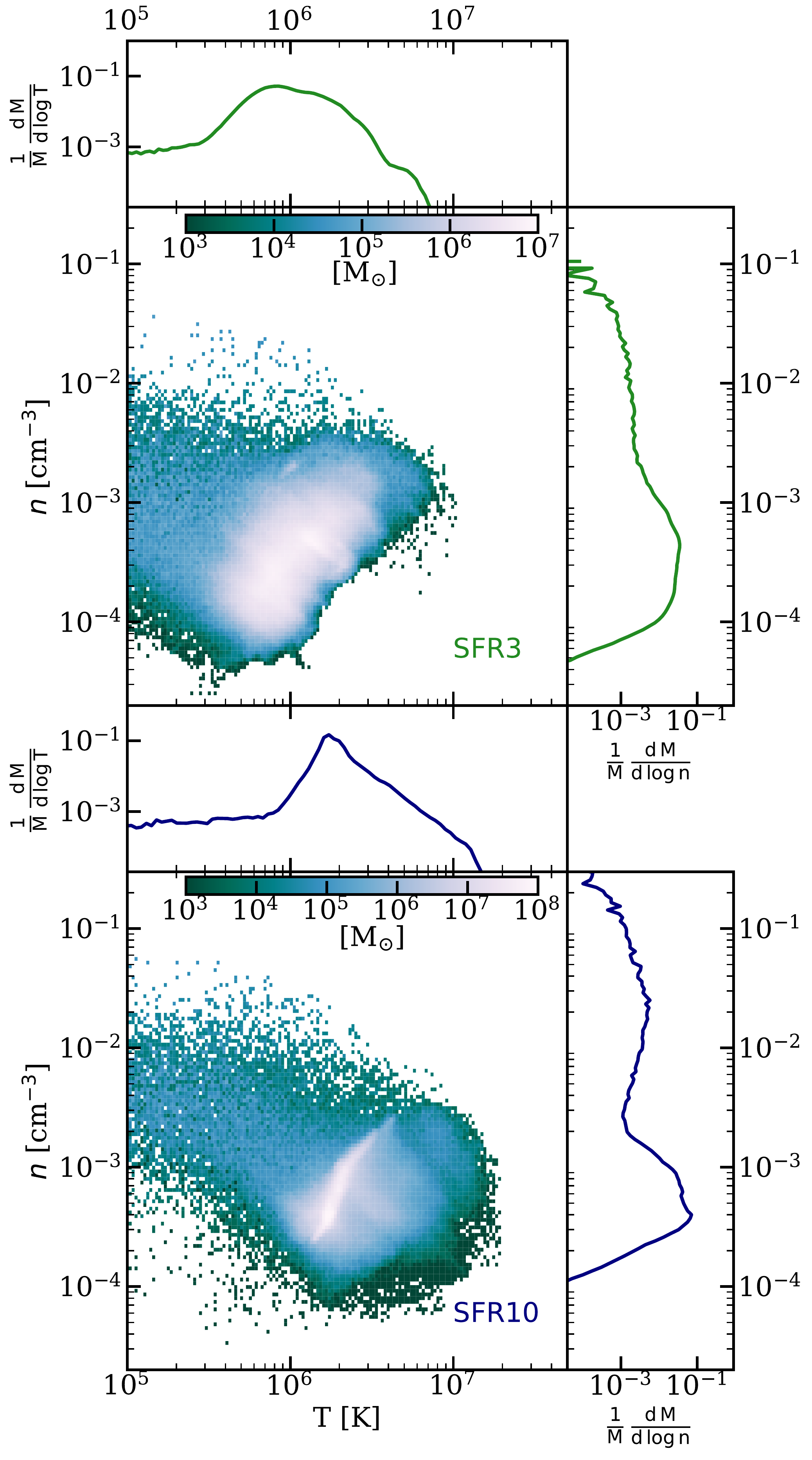}
    \caption{Mass distribution of density and temperature distribution for \lsfr\, (top) and \hsfr\, (bottom) for the region of interest ( essentially a sphere of radius $50$ kpc from which we remove a cylinder, $\pi 20\times20\times (\pm 3$) kpc$^3$). The snapshots are at t=$3$ and $1.3$ Gyr, respectively. The colour shows mass in each density and temperature bin. The side panels show mass distribution of density (right) and temperature (top). Both runs show that hot gas has a broad distribution of density and temperature. }
    \label{fig:denT-histo}
\end{figure}

\section{Simulations}\label{sec:simulations}

In this Section, we briefly discuss the simulation setup of and results from L20, data from which are used for analysis for this paper. L20 uses 3D hydrodynamical simulations to investigate how supernovae (SNe)-driven hot outflows regulate the CGM around a Milky Way-mass galaxy. The outflow model -- in particular, the outflow rate of mass, energy, and metals -- is taken from  kpc-box simulations with pc-scale resolutions that model the launching of outflows by SNe for various star formation surface densities \citep{Li17a}. The hot outflow properties show a very good agreement among small-box simulations of supernovae-driven outflows which utilized different hydro-codes \citep{li_bryan20}. Outflows driven from the ISM is multiphase, and the hot phase has $T \sim 10^{6-7}$ K. For L20, only the hot phase of the outflows is added from small-box simulations because the hot phase is much more powerful than the cool phase, and can travel to distances $\gtrsim 100$ kpc away from the galaxy (see discussions in \citep{li_bryan20}). In contrast, most cool outflows can only travel to a few kpc \citep{Li17a,Kim&Ostriker17,Vijayan+18,li_bryan20}.

For a detailed description of the simulation setup, one is referred to L20. Here is a brief summary: 
The simulation domain is [$800$ kpc]$^3$ containing the galaxy at its centre. The galaxy disk in the $x-y$ plane. The gravitational potential is determined by a \cite{Burkett95} profile for the dark matter (DM) halo of $10^{12} M_\odot$, a Plummer-Kuzmin profile \citep{Miyamoto-Nagai75} for stellar disk and a spherical \cite{Hernquist93} profile for the bulge. The spatial resolution is progressively finer near the center, with $0.39$ kpc in the inner [$50$ kpc]$^3$, $0.78$ kpc in the [$100$ kpc]$^3$, and so forth. The Grackle library \citep{Smith+17} is used to calculate the metallicity-dependent gas cooling. The reader is directed to Section $3.1$ of L20 for further details on cooling.

The initial gas distribution is in overall hydrostatic equilibrium with the DM potential with an inner core at radii $<$ $40$ kpc (see Figure $1$ of L20 and is overwritten by subsequent gas evolution. The initial gas distribution does not include cold gas as gas at this temperature has minuscule contribution to the X-ray emission, which is the primary focus of this paper. The gas has a uniform temperature of \mk\, i.e. around the virial temperature of Milky Way (MW) halo, and a low metallicity $Z = 0.2$ \zsolar, where $Z_\odot\equiv 0.01295$. The total gas mass is a free parameter, represented by n$_0$, the mean number density within $R=200$ kpc. The fiducial $n_0$ is determined by comparing $L_X$ of the resulted CGM to the observations with the same stellar mass and star formation rate (SFR). 

The outflows are parameterised using the mass, metal and energy loading factors multiplied by the designated SFR (see Equations 7, 8 and 9 of L20).
The outflows from star formation episodes are injected in a small region above and below the disk at a height of 3 kpc (no cool gaseous disk is present in the simulation as initially). Temporally, each star formation event occurs every $\Delta t \simeq 9.9$ Myr with a randomly chosen $x-y$ location. The added outflows are metal-enriched and initially overpressured, and they expand subsequently and interact with the ambient gas.

In this paper, we analyse the results from two simulations with different parameterizations. One is when star formation (SF) is relatively quiescent and widespread on the disk, like the current state of the MW. The other is where SF is more intense and centrally concentrated. The different cases have SFRs of $3$ and $10$ \sfr, and $\dot{\Sigma}_{\rm SFR}$ of 6 $\times 10^{-3}$ and 0.3 M$_\odot$ kpc$^{-2}$ yr$^{-1}$, respectively. From the small-box simulations, such $\dot{\Sigma}_{\rm SFR}$ correspond to mass, energy, and metal loading factors of hot outflows $[1.0, 0.3, 0.5]$ and $[0.2, 1.0, 0.5]$, respectively. 
In the former case, the specific energy of hot outflows is lower and outflows cannot escape from the DM halo of $10^{12} M_\odot$ , therefore they form a large-scale fountain around the galaxy. In the latter, hot outflows are powerful enough to escape the halo and form a bipolar structure (Li \& Tonnesen, in prep). In both cases, cool CGM gas precipitates out of the hot atmosphere at later times. We denote the two different runs as \lsfr\ and \hsfr. The X-ray luminosity $L_X$(0.5-2.0 keV) of the CGM for the two cases are 8$\times 10^{39}$ and $2\times 10^{40}$ erg s$^{-1}$, respectively. 

Most of the soft $L_X$ of the CGM comes from $\lesssim[30$ kpc]$^3$, where density is highest (as shown in Figure \ref{fig:projected-emissivity} below). Consequently, our analysis is limited to the inner [$50$ kpc]$^3$ of the simulation box. 
We analyse the simulation data for these runs at times after $L_X$ has reached a steady state. We show the time evolution of the results in Appendix \ref{app:temporal_variation_mass_dist}. In Section \ref{sec:discussion}, we briefly discuss our results in the context of other simulations.

\section{RESULTS}\label{sec:results}

\subsection{Visualization of Simulated CGM}\label{subsec:visualization_of_simulations}

We first discuss the distribution of various physical quantities within a radius of $50$ kpc of the simulation box.

In Figure \ref{fig:x-slices}, we show $x=0$ slices (galaxy disk viewed edge-on) of density, temperature, metallicity and radial-velocity for \lsfr\ (top) and \hsfr\ (bottom). The snapshots are taken at $3$ Gyr and $1.3$ Gyr for \lsfr\ and \hsfr, respectively. At these time steps the outflows have significantly changed the initial conditions of the halo gas and the CGM has reached a quasi-steady state (see Appendix \ref{app:temporal_variation_mass_dist} for a discussion).

Fast and hot outflows emerge from each star formation episode, disrupting the initial isothermal gas profile. In \lsfr\ these episodes result in formation of expanding bubbles, distinctly visible as nearly spherical regions with low densities and high temperatures (left two panels on top). 

In \lsfr, at smaller radii (R$\lesssim 10$ kpc), low-density, high-temperature bubbles expand into the surrounding CGM. Every outflow injection event drives an outward-moving shock into the ambient medium. Beyond the outer-shock, the density and temperature profiles decrease with radius. Gas which cools at large radii loses pressure support and falls under gravity. We can clearly see this cooled gas as a high-density ($\sim 1$ \cc) and low-temperature feature ($\sim 10^4$ K) feature in the top panel of Figure \ref{fig:x-slices}. Eventually, this gas settles at the bottom of the potential well, that is, at the centre of the box. 
Due to a long period of star formation, the metallicity of the ambient medium has been completely overwritten by gas with a higher metallicity around $\sim 1.2$ \zsolar. The spherical outflows have expanded in all the directions and produced uniform metallicity in the region. The outflowing gas has velocities of a few $\sim 100$s \kmps. For \lsfr, the hot outflows do not possess sufficiently large velocities. At large radii, these outflows turn around and form large-scale fountain flows.

Outflows in \hsfr\ have a bi-conical shape because they have a specific energy larger than the halo potential, and are able to break out from the halo. At the launching site, the outflows are hotter than the ambient medium. As the injected gas progresses into the CGM, it mixes with the pre-existing hot gas, creating a wide distribution in density as well as in temperature.\footnote{Note that such a mixing process between ``hot'' and ``hotter'' material is different from mixing occurring between 10$^4$ and 10$^6$ K, which arises when cold gas is entrained by hot outflows and produces cooler, mainly UV-emitting gas \citep{Suchkov+96, Fielding&Bryan21, Nguyen+21}.} The cold and dense gas seen in the slices is produced by the precipitation of the CGM as a result of the interaction between hot outflows and pre-existing halo gas. The gas within the $50$ kpc radius show metallicity values over a wide range. The most enriched part, at $\sim 4$ \zsolar, is the newly injected hot outflows. Lower-metallicity gas is created as a result of mixing between these super-\zsolar outflows and the low-metallicity ambient medium. 
In the bipolar region, velocities are large and predominantly outward, showing that outflows are powerful enough to break out through the pre-existing gas.

\subsection{Histograms of Density and Temperature}\label{subsec:rho-T-histos}

The temperature and density of the hot gas are critical in determining the X-ray spectra. In this Section, we will discuss in detail the density and temperature distribution of the simulated CGM.

For the analysis of hot gas, we have removed the ``galaxy'' region, from a sphere of radius $50$ kpc, due to the proximity to the injection region. The removed region is a cylinder with its axis along the $z-$axis, $\pi 20\times20\times (\pm 3$) kpc$^3$. In X-ray observations of diffuse gas in CGM, the region near the disk is also removed in order to include only contribution from the extra-planar region and to remove contribution from point X-ray sources. We will refer to the remaining region, i.e. total box minus the galaxy, as the ``region of interest''. 

In Figure \ref{fig:denT-histo}, we show the mass distribution of density and temperature for the regions of interest of the two runs. They are taken at the same snapshots as Figure \ref{fig:x-slices}. The colour indicates mass per density and temperature bin. It is clear that in both runs, the gas occupies a wide range in density and temperature. 

The side panels of Figure \ref{fig:denT-histo} show 1-D fractional mass distribution of density (right) and temperature (top) of the gas in the region of interest. Despite morphological differences in the two runs (see Figure \ref{fig:x-slices}), the 1-D profiles are qualitatively similar-- the X-ray emitting gas shows broad distributions around a single peak. The temperature profile is smoother and broader for \lsfr. In \hsfr, the peak of the 1-D temperature distribution is sharper. Around this peak is the gas temperature of the ambient medium outside the bi-conical region, which has not been largely affected by the outflows.

The mass distribution presented in this section represents a snapshot in the simulation. However, we note that these times are after the systems have reached a quasi-steady state and the distributions do not change thereafter. We show the temporal variations in the mass distribution in the Appendix \ref{app:temporal_variation_mass_dist}.

\subsection{X-ray Emission}\label{subsec:X-ray_emission}

\begin{figure}
	\includegraphics[width=\columnwidth]{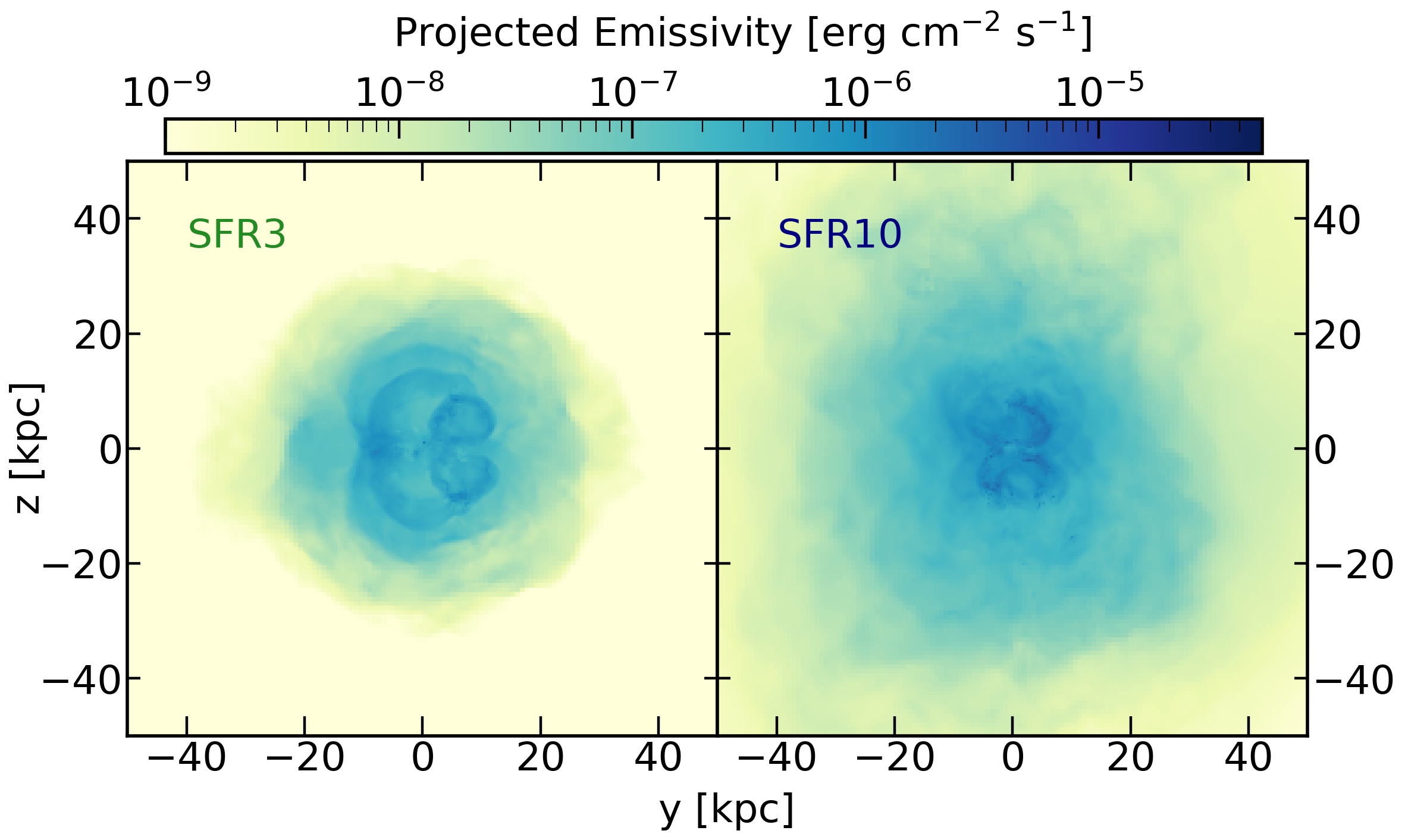}
    \caption{Projected emissivity at $0.5-2.0$ keV for \lsfr\, (left) and \hsfr\, (right) runs.In \lsfr\ case, emission is most significant for radii $\lesssim 20$ kpc, while for \hsfr\ the emission is more extended. The bi-conical structure of \hsfr\ is not very prominent in the projected emission map because the emission is dominated by gas which lies outside the outflow cones.
    }
    \label{fig:projected-emissivity}
\end{figure}

\begin{figure}
	\includegraphics[width=\columnwidth]{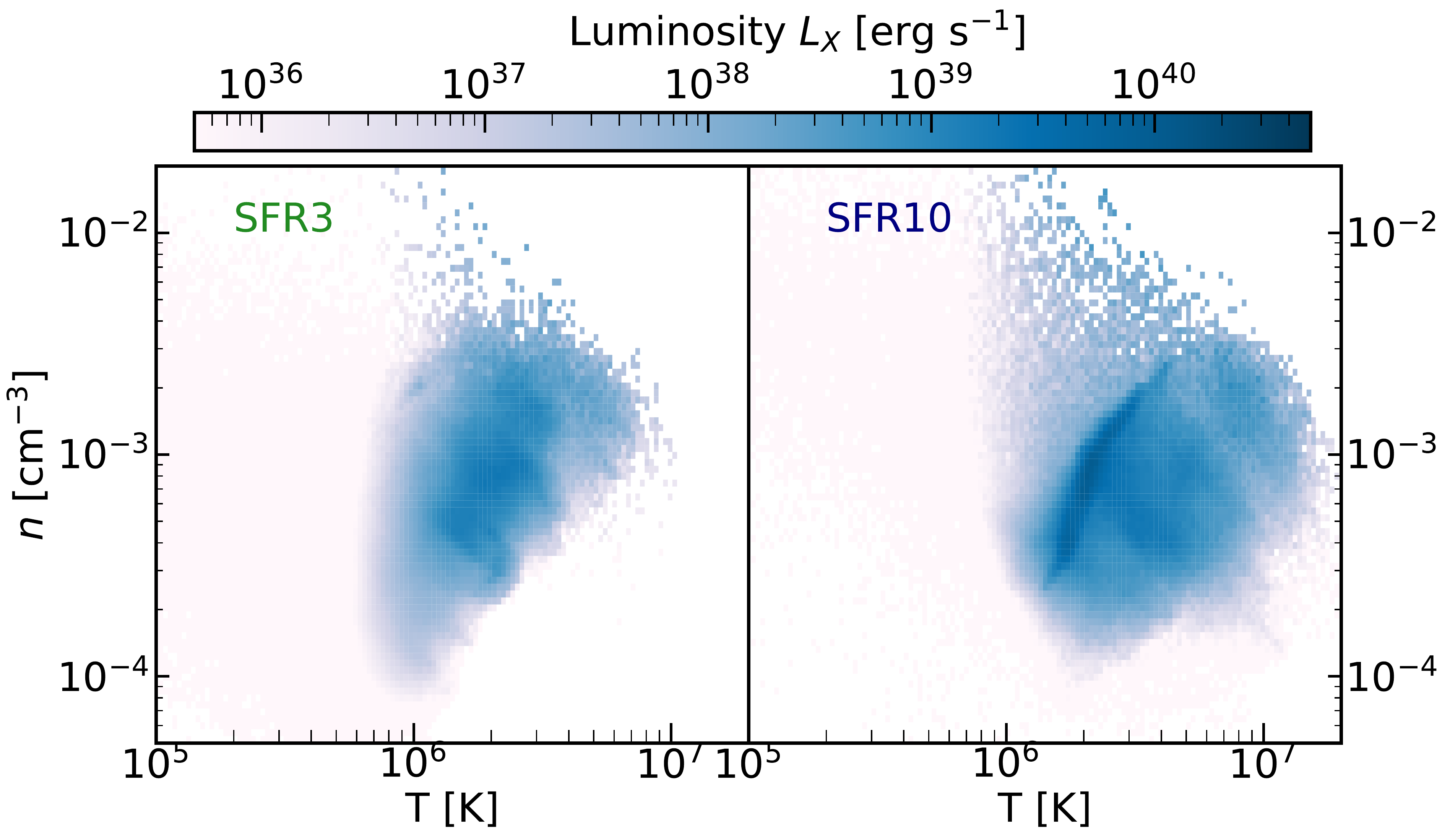}
    \caption{X-ray luminosity $L_X$ as a function of density and temperature. This is similar to Figure \ref{fig:denT-histo} but the colour bar represents $L_X$ ($0.5$-$2.0$ keV). The luminosity distribution closely follows the mass distribution as shown in Figure \ref{fig:denT-histo}.}
    \label{fig:luminosity-histo}
\end{figure}

Figure \ref{fig:projected-emissivity} shows the edge-on view of the projected emissivity at $0.5-2$ keV for the two runs. We obtain X-ray emissivity using the \textit{add\_xray\_emissivity\_field} module of \texttt{yt} with the APEC emission tables \footnote{APEC v 2.0.2 downloaded from \hyperlink{https://yt-project.org/data/}{https://yt-project.org/data/}.}. Firstly, \hsfr\ has a more extensive and brighter emission than \lsfr. For \lsfr, the high-emissivity regions lie along the shells corresponding to recent outbursts. Though the hot outflows have expanded to fill the entire region of interest, the gas beyond the radius of $\sim 20$ kpc has lower densities as well as lower temperatures and does not emit as strongly in X-rays. In \hsfr, the hot ($\gtrsim 10^6$ K) gas also fills the entire region of interest. Though, the bi-conical feature of the outflows is not very prominent on the emissivity map because the emission is dominated by the gas lying outside of the cone.

In Figure \ref{fig:luminosity-histo}, we show the distribution of X-ray luminosity $L_X$ ($0.5-2.0$ keV) as a function of density and temperature for the regions of interest. The distribution of X-ray luminosity closely follows that of the mass, as shown in Figure \ref{fig:denT-histo}. This is not surprising because luminosity is proportional to the gas mass. Gas with densities of $10^{-4}$ -- $10^{-3}$ \cc and temperatures of one to a few $10^6$ K dominates the X-ray emission. For \lsfr, the emission is dominated by gas with temperature $\gtrsim 10^6$ K, even though significant mass lies near the peak of the cooling curve, i.e., $10^{5-6}$ K. This is expected as $10^{5-6}$ K gas emits mostly in UV bands.

\subsubsection{Spectra Generation}\label{subsec:spec-generation}

In this Section, we describe how we construct the spectra from the simulated CGM. We use the python-based module pyXSIM\footnote{ \hyperlink{http://hea-www.cfa.harvard.edu/~jzuhone/pyxsim/index.html}{http://hea-www.cfa.harvard.edu/~jzuhone/pyxsim/index.html}.}, an implementation of PHOX \citep{Biffi+12, Biffi+13}. pyXSIM generates X-ray photons in each cell of the simulation domain using the gas density, temperature and metallicity and an X-ray emission model based on the APEC tables which assume collisional ionization equilibrium (CIE)\footnote{The timescale to achieve CIE is $\sim 10^4$ yr, which is much shorter than both dynamical and cooling time scales of the gas.}. The total number of photons generated also depends on input parameters such as exposure time ($100$ ks), effective area ($1000$ cm$^2$), and the distance of the source ($1$ Mpc). We have chosen arbitrarily large values for effective area and exposure time in order to generate sufficient number of photons for the next steps. The coordinates and distance of the source for the spectra discussed in this Section are also arbitrary.
From this collection of photons, a subset is projected along the line-of-sight and a Galactic absorption model is applied, assuming a hydrogen column density of $1.93\times 10^{20}$ cm$^{-2}$. We use the Wisconsin Absorption Model \citep[``wabs",][]{Morrison1983} and generate spectra in the X-ray range $0.5-2.0$ keV. We note that the spectra shown are not convoluted with any instrument response. 

Using this procedure we produce the spectra for \lsfr\ and \hsfr\ and present them in the topmost and the fifth panels of Figure \ref{fig:spectra-all}, respectively. In the left panel we show the low-resolution spectra which is binned at $130$ eV\footnote{Corresponds to the resolution of Chandra ACIS in the wide field imaging mode- https://cxc.harvard.edu/cal/Acis/.}. 
The right panel shows the high-resolution spectra, binned at $1$ eV. The differences in the temperature distribution in the two runs are reflected in their respective spectra. Because of the lower temperature of the gas present in \lsfr, its spectra is steeper than that of \hsfr.

\subsection{1-T and 2-T Models}\label{subsec:1T-2T-spec}

\begin{table}
\caption{Best-fit parameters for the simulated spectra}\label{tab:spectra-params}
    \begin{tabular}{ | p{2.5cm} | p{2.25cm} | p{2.25cm}|} 
    \hline
Parameter     & \lsfr & \hsfr \\ \hline
T$_{\rm 1T}$ [MK]            & 2.2  & 3.8  \\ \hline
T$_{\rm 2T,low}$ [MK]        & 0.8  & 1.7  \\ \hline
T$_{\rm 2T,high}$ [MK]       & 2.8  & 3.0  \\ \hline
Z [Z$_{\odot}$]              & 1.19         & 0.58         \\ \hline
$\mu{_T}$ [MK]            & 1.0          & 2.2          \\ \hline
log($\sigma_{\rm T}$)     & 0.51          & 0.46         \\ \hline

\end{tabular}
\end{table}

In this Section we introduce the 1-T and 2-T models for fitting X-ray spectra from observations. We will then compare the spectra generated using these models with those generated using the simulation data.

The 1-T and 2-T models of plasma emission are commonly used to fit the observed X-ray spectra. The 1-T model assumes that the X-ray emitting gas is at a single temperature, while the 2-T model assumes that the observed spectra is results from a superposition of spectra at two different temperatures.

\subsubsection{Spectra from 1-T Model}\label{subsubsec:1T-spectra}

To construct the 1-T model, we generate a ``uniform box'', a \texttt{yt} object having identical density and temperature values in all cells. To keep the model simple, we set its metallicity as the average metallicity of the gas in the simulation. To get the best fit parameters for temperature, we adopt a ``blind-search'' method. We use the spectra from simulation as the ``ground truth'' and generate multiple spectra with varying temperatures of the uniform box. To determine the best-fit spectrum for a run, for each high resolution spectrum from the uniform box we calculate the net fractional difference (NFD) which is given by $\sqrt{\rm{\Sigma\Big((Simulation - Model)}^2/\rm{Simulation}^2\Big)}$. 

We note that the quantity we define as NFD is different from the conventional ``reduced chi square'', the standard statistic used by observers to determine the goodness of the fit, given by $\chi_{\rm reduced}^2\equiv \Sigma_i \frac{(O_i-C_i)^2}{\nu \sigma^2_i}$, where $O_i$, $C_i$, $\sigma_i^2$ and $\nu$ are the observations, the model and the variance, and the degrees of freedom of the data, respectively. We stress here that for our purpose the NFD defined by us suits well. We select the temperature for which the NFD is minimum, indicated in Table \ref{tab:spectra-params}. Once the best-fit temperature is set, we adjust the normalisation of the spectra to minimise the NFD again. The values of NFD for the best-fit temperature and normalisations are indicated in Figure \ref{fig:chi}.

We represent the best-fit 1-T spectra using $\Delta$, which is the fractional difference between the simulation spectra and the best-fit spectra at a given energy. $\Delta$ is given by $\Delta$=(Simulation - Model)/Simulation and it characterises the deviation of the model from the simulation. The second and sixth panels of Figure \ref{fig:spectra-all} show the $\Delta$ for 1-T model for \lsfr\ and \hsfr, respectively. 
For \lsfr, $\Delta$ becomes negative in the higher energy range as the model over-predicts emission from the CGM. Because the model is biased towards higher temperatures, for \hsfr\ , $\Delta$ is large in the lower energy range, while it is close to zero for higher energies. 
Despite its simplicity, the spectra from the 1-T model fits the simulation spectra well for both the runs. The models fits the spectra especially well for high energy range ($\gtrsim 1.0$ keV). We note that for low-resolution spectra (left panels) it appears that 1-T model is a close fit. 

In Figure \ref{fig:1T-mass-temperature}, we depict a visualization of the 1-T model. In top and bottom panels, the solid curves show the fractional mass distribution of the logarithmic temperature for the simulations, which are identical to the top panels of Figure \ref{fig:denT-histo}. The best-fit 1-T models are represented by dashed vertical line in the Figure.

From Figure \ref{fig:spectra-all}, we conclude that 1-T model is able to match the spectra well. However, the mass distribution represented by the 1-T model is drastically different from the underlying temperature distribution of the gas, which is quite broad. Thus, we conclude that a 1-T model, while being able to reproduce the spectra, cannot faithfully represent the physical temperatures in the CGM.

\begin{landscape}\centering
\begin{figure}
    \includegraphics[width=1.3\textwidth]{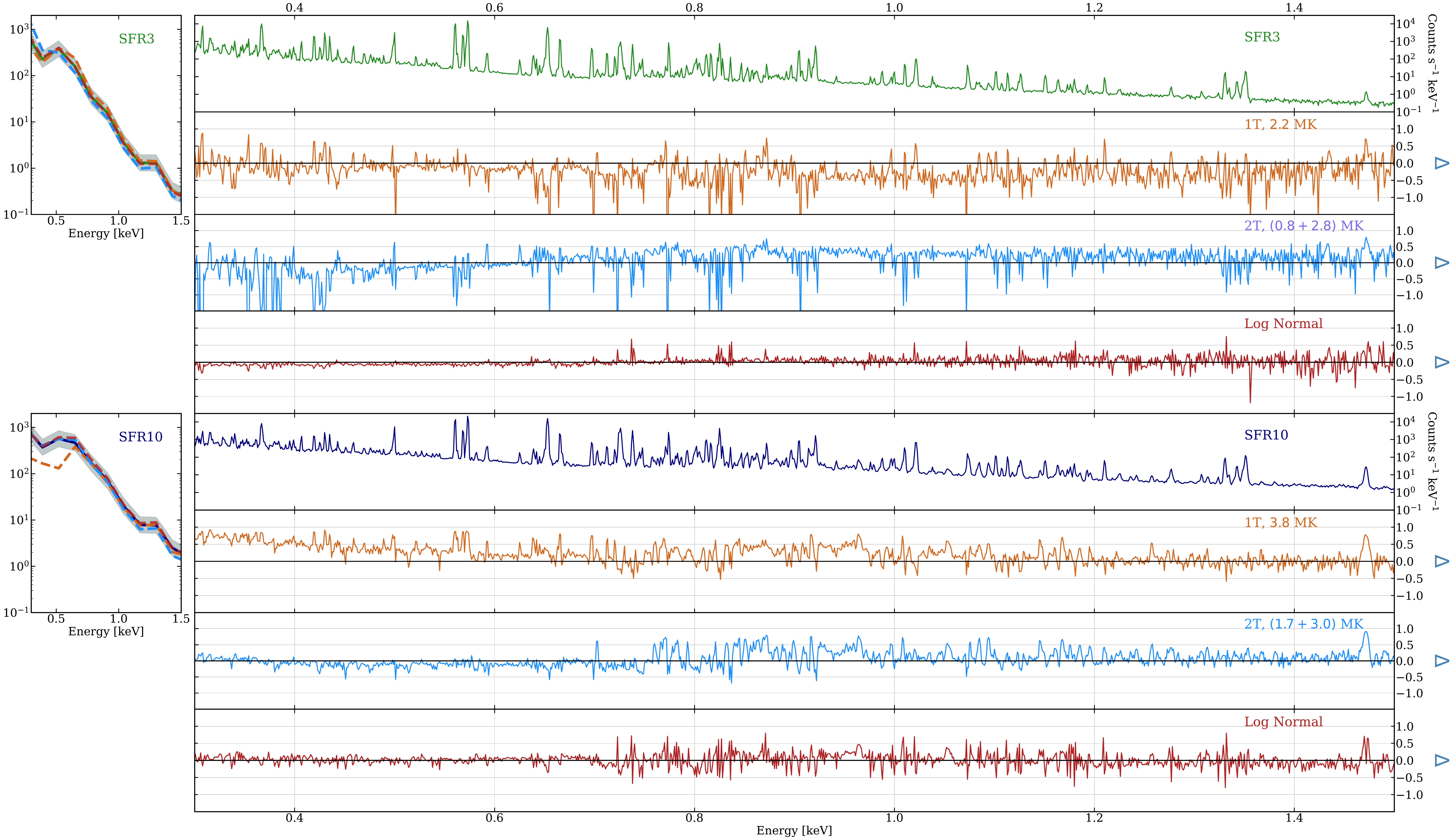}
    \caption{
    Spectra from simulations compared with best-fit spectra from different models. Spectra on the left are binned at $130$ eV, similar to the Chandra resolution, while those on the right are binned at $1$ eV. Solid lines represent the spectra from the simulations of \lsfr\ (green) and \hsfr\ (blue) at $3.0$ Gyr and $1.3$ Gyr respectively. In the left column, the best-fit spectra from 1-T, 2-T and the log-normal models are represented by orange, blue and red curves, respectively. In the right column panels, we compare the high resolution spectra from simulations with the corresponding $\Delta$ values from the different models- 1-T (orange curves in second and sixth panels), 2T (blue curves in third panel from top and second panel from the bottom), and log-normal (red curves in fourth panel from top and the last panel). The grey shaded region in the left panels indicates the temporal variations in the spectra over the duration of the simulation. 
    }
    
    \label{fig:spectra-all}
\end{figure}

\end{landscape}

\begin{figure}
\center
    \includegraphics[width=\columnwidth]{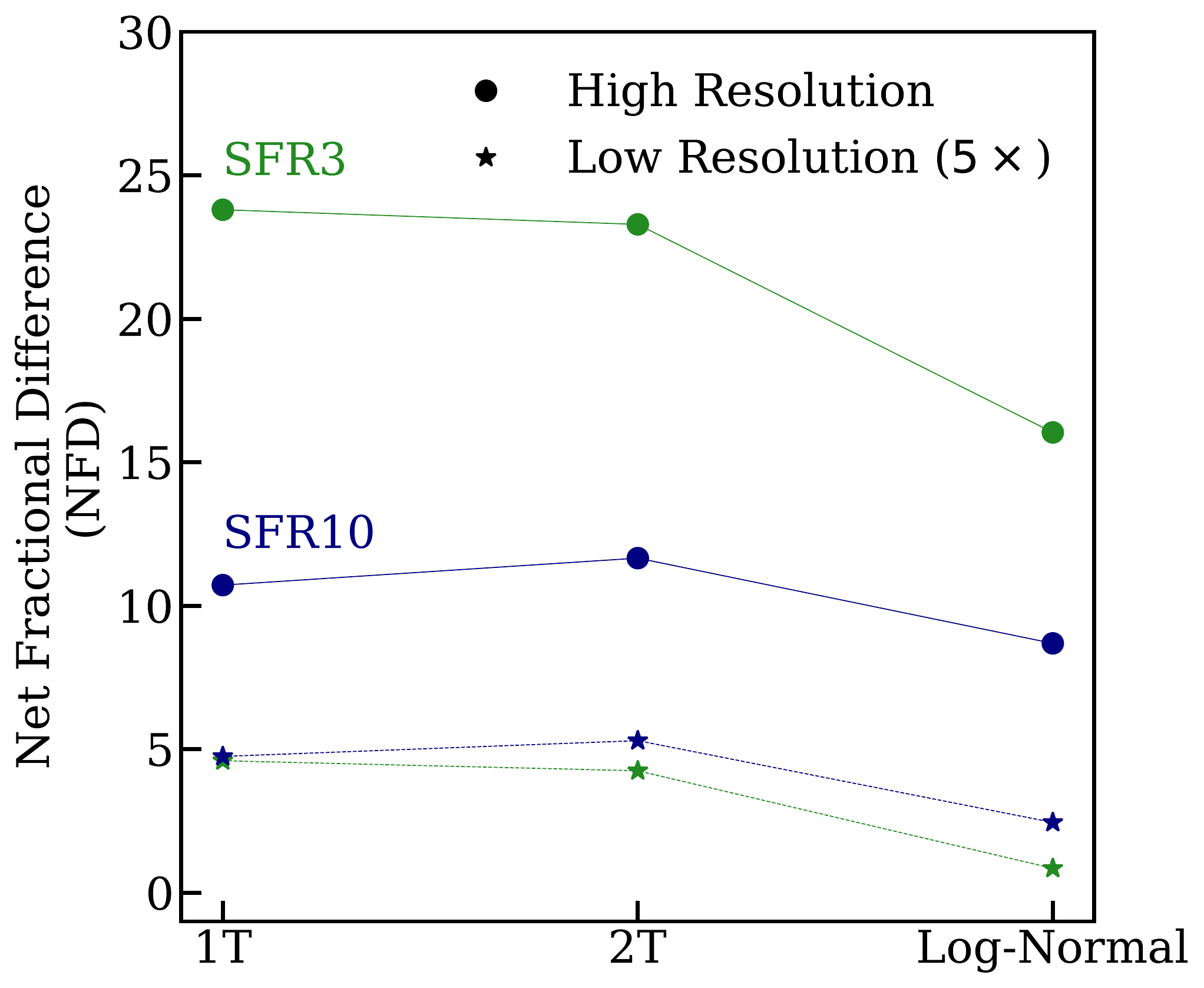}
    \caption{Comparison of NFD $\equiv \sqrt{\Sigma \Big(\rm{(Simulation - Model)}^2/\rm{Simulation}^2\Big)}$ among 1-T, 2-T and Log-Normal models. Circles and stars represent \lsfr\ and \hsfr, respectively. We have multiplied the NFD values for the low-resolution spectra by a factor of $5$ for clarity. The log-normal model leads to a better fit (smaller NFD than the 1-T/2-T models. We point out here that the number of bins used for high- and low- resolution spectra are different and thus, the curves for high- and low-resolutions should not be compared.}
    \label{fig:chi}
\end{figure}

\begin{figure}
\center
    \includegraphics[width=\columnwidth]{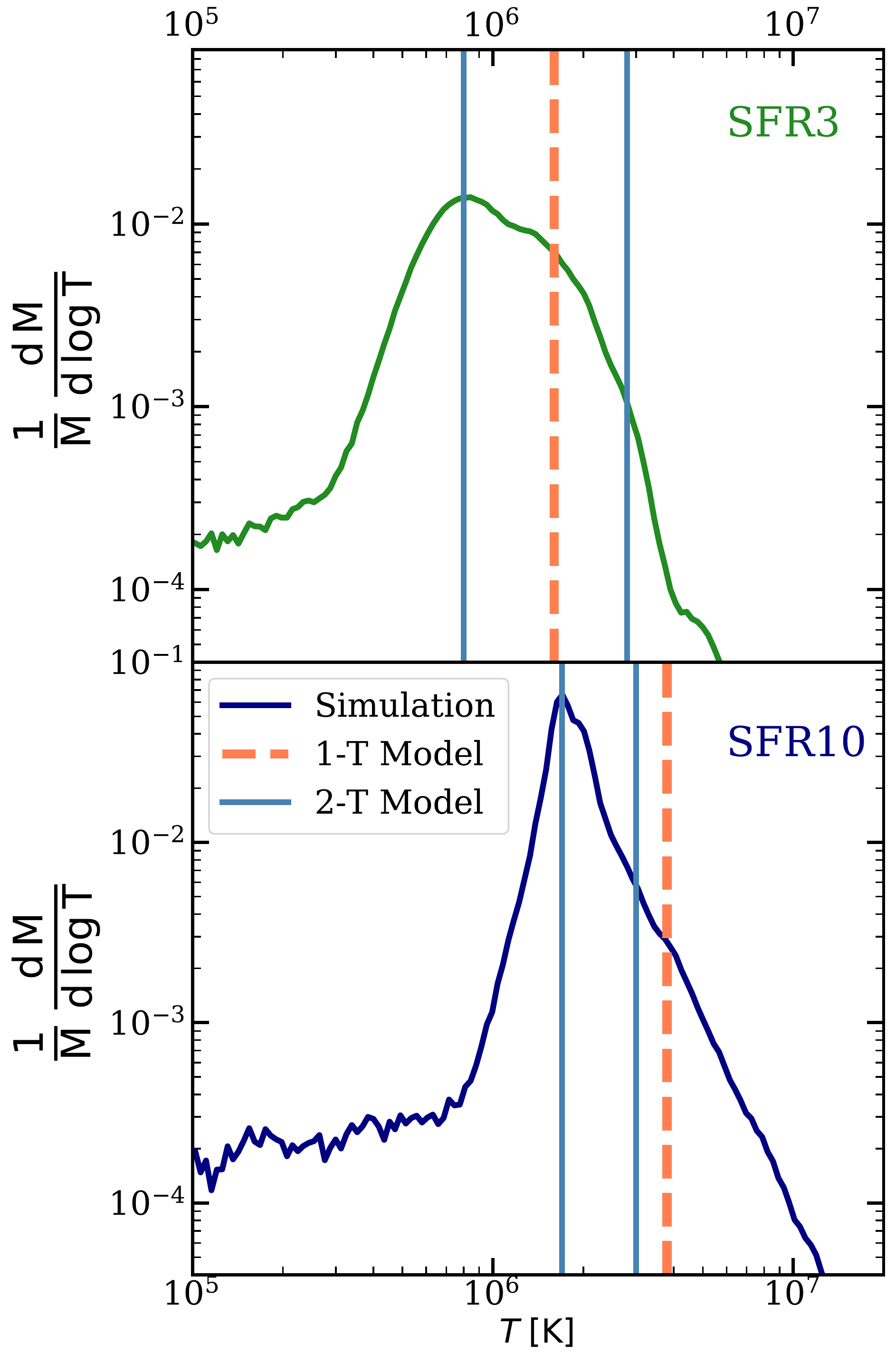}
    \caption{Comparison of the 
    mass distribution of the temperature between simulations (solid curves) and best-fit parameters for the 1-T/2-T models (vertical lines). 
    The lower-temperature of the 2-T model is set by hand to be $T_{\rm peak}$. The 1-T/2-T model cannot reflect the true temperature distribution of the gas.
    }
    \label{fig:1T-mass-temperature}
\end{figure}

\subsubsection{Spectra from 2-T Model}\label{subsubsec:2T-spectra}

When fitting for the observed data, if a 1-T model cannot reproduce the spectra, a 2-T model is usually invoked, which adds an additional temperature component.
Three parameters determine the shape of the spectra in the 2-T model, viz., the two temperatures and the ratio of their normalisations. However, it is possible that more than one combinations of these parameters can fit the same data reasonably well. Instead of exploring a full parameter space, we here discuss one possible way of finding the best-fits for these parameters for the simulation spectra. Our procedure for fitting the spectra differs from the way spectra are fit by observers. While we fix one temperature and iterate to find the second temperature, traditionally observers pursue to fit the two temperatures simultaneously and minimize the standard reduced chi-squared statistic.

Following the method described in Section \ref{subsubsec:1T-spectra} for the 1-T model, we now construct two separate uniform boxes, each representing one of the two temperatures of the 2-T model. For simplicity, we use T$_{\rm peak}$, the temperature at which the mass-weighted temperature distributions (see Figure \ref{fig:denT-histo}) peak, as the lower temperature for this model denoted as T$_{\rm 2T,low}$. Next, starting from a guess value for the higher temperature, T$_{\rm 2T,high}$, we calculate NFD for a range of temperatures. As previously, we obtain the best match to the simulations spectra by selecting the T$_{\rm 2T,high}$ with the smallest NFD. We indicate T$_{\rm 2T,low}$ and T$_{\rm 2T,high}$ for the two runs in Table \ref{tab:spectra-params}. We then adjust the normalisations of the two uniform boxes to minimize NFD. 

We obtain the spectra from the uniform boxes at T$_{\rm 2T,low}$ and T$_{\rm 2T,high}$ and the total 2-T spectra is obtained by summing their normalised values. In Figure \ref{fig:spectra-all}, we show the 2-T spectra binned at at $130$ eV (left) and $1$ eV (right). As previously, we show how $\Delta$ between the simulated and the best-fit 2-T spectra in the third from top and second from bottom for \lsfr\ and \hsfr, respectively. 
The difference between the model and the simulation spectra is within a factor of $2$. As for the 1-T model, $\Delta$ is noisier for \lsfr\ as compared to \hsfr. From Figure \ref{fig:chi}, we see that NFD for the 2-T model is nearly same as that for the 1-T model.

In Figure \ref{fig:1T-mass-temperature} the 2-T model is represented by two vertical solid lines at T$_{\rm 2T,low}$ and T$_{\rm 2T,high}$. Same as for the 1-T model, though the modelled spectra match the simulation spectra well, the temperature distributions from 2T models deviate far from the actual distribution. Further, the best-fit T$_{\rm 2T,high}$ has little physical meaning. 

\section{Log Normal }\label{sec:log-normal}

In the previous section, we discussed the 1-T and the 2-T models, which assume that all of the mass emitting X-ray has either one or two temperatures. Though such models are able to match the simulation spectra reasonably well, the temperature distributions in these models are not physically motivated. In this Section, we propose a novel temperature model for fitting with the spectra.

As discussed in Section \ref{subsec:rho-T-histos}, the temperature and density distributions in the simulations cover a broad range around a single peak. We, therefore, model the temperature distribution  with a log-normal distribution, which is expressed mathematically as

\begin{equation}
    p(T) = \frac{1}{\sigma_{T} T \sqrt{2\pi}} e^{-\frac{(\rm{ln}\ T - \mu_{T})^2}{2\sigma_{T}^2}},
\label{eqn:log-normal}
\end{equation}
where $\sigma_{T}$ and $\mu_{T}$ are the standard deviation and the mean of the underlying log-normal distribution, respectively.

We generate a ``log-normal'' box such that the mass distribution in temperature within the box follows a log-normal distribution. To generate a temperature in the log-normal box, we begin with a guess value for $\sigma_{\rm T}$ and $\mu_{\rm T}$, and then generate distributions for gas temperature using Equation \ref{eqn:log-normal}. The density values of the gas in this box also follow a log-normal distribution similar to Equation \ref{eqn:log-normal}.
We generate the spectra and compare it with the simulation spectra. We then adjust the values of these parameters, i.e., $\sigma_{\rm T}$ and $\mu_{\rm T}$, to find the best-fit parameters in few iterations with minimized NFD. Once the best-fit temperature value has been obtained, we adjust the mass of gas in the log-normal so as to get the minimum NFD. The best-fit values of $\sigma_{\rm{T}}$ and $\mu_{\rm T}$ are listed in Table \ref{tab:spectra-params}.

The fourth and the last panels of Figure \ref{fig:spectra-all} show the $\Delta$ between the simulation spectra from \lsfr\ and \hsfr, respectively.
We show both the low-resolution (left) and the high-resolution (right) spectra.
We note that the log-normal spectra match the simulations spectra well at both high- and low-energy ends. We indicate the NFD values for this model in Figure \ref{fig:chi}. For both \lsfr\ and \hsfr, these values are lower than the corresponding values for 1-T and 2-T. This indicates that the log-normal model yields a better fit than the 1-T/2-T models.

\begin{figure}
\center
    \includegraphics[width=\columnwidth]{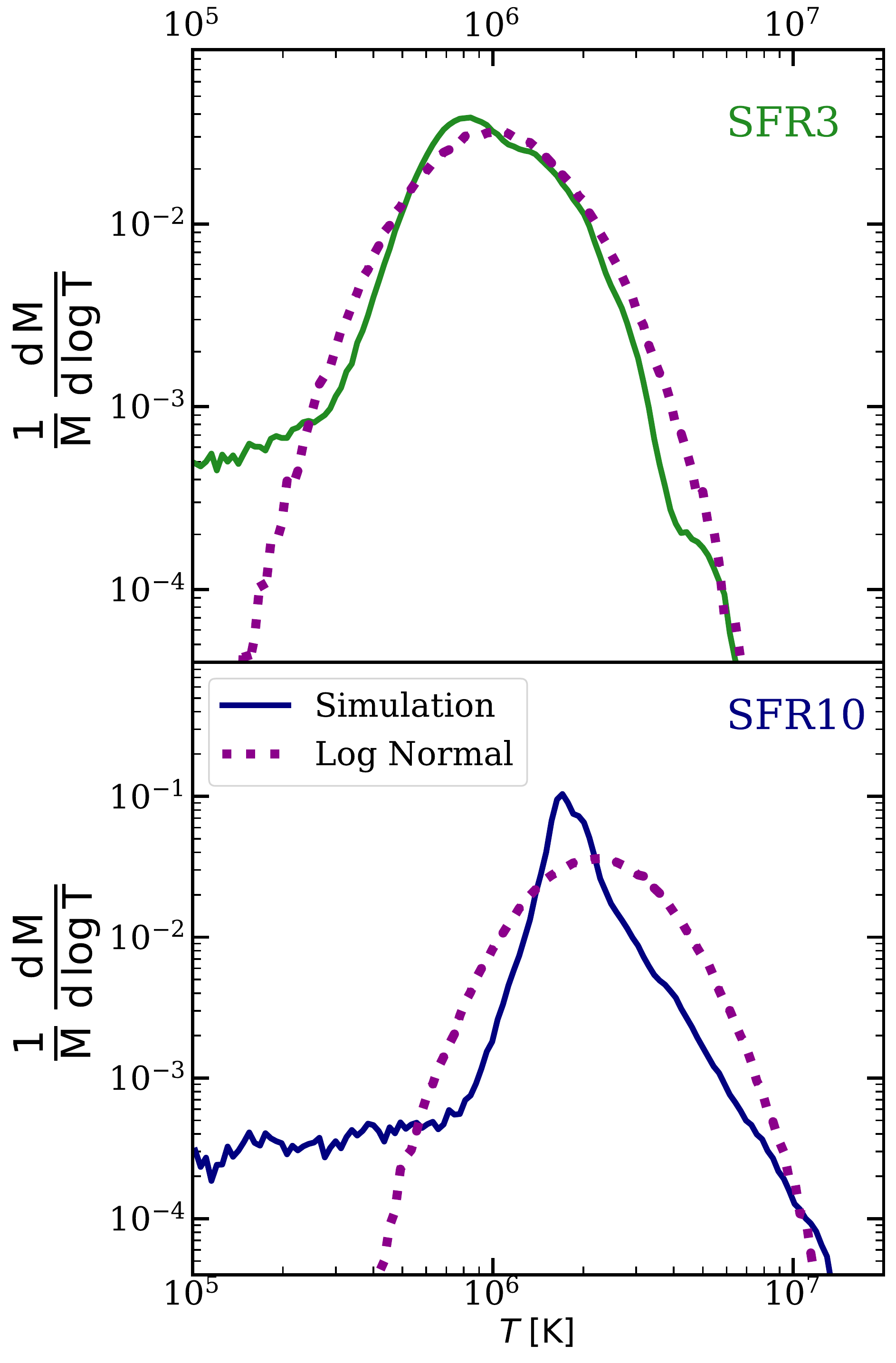}
    \caption{Comparison of the mass distribution of the temperature between simulations (solid curves) and the best-fit log-normal model (dotted curves). The shape of log-normal model better reflects the broad temperature distribution of the gas.
    }
    \label{fig:ln-td}
\end{figure}

Figure \ref{fig:ln-td} shows the comparison of the fractional mass distribution. The solid curves indicate the mass distributions from simulation and the dotted curves represent the best-fit for high resolution spectra. The log-normal distribution reproduces the broad distribution in temperature, unlike the vertical delta function used in 1-T/2-T models. The models are able to generate mass fractions at both low and high temperatures. Admittedly, the mass distribution for \lsfr\ is a better fit than that for \hsfr. The reason for this difference in fits is that outflows in \hsfr\ leave the gas outside of the bi-cone relatively undisturbed; the undisturbed ambient medium contributes the sharp peak at $\sim 2\times 10^6$ K to the mass distribution in Figure \ref{fig:ln-td}. We discuss the consequences of bi-polar outflows on spectrum of CGM gas in greater detail in the next Section. That said, the essence of a broad T-distribution is captured for both cases by log-normal model. The fact that the log-normal model can reproduce both the spectra and underlying T-distribution demonstrates its feasibility as a physically-based model, which we argue is better than the 1-T/2-T models.

\section{Spectra from Inside Versus Outside of Outflow Bi-cones}\label{sec:perp-spec}

\begin{figure}
    \includegraphics[width=\columnwidth]{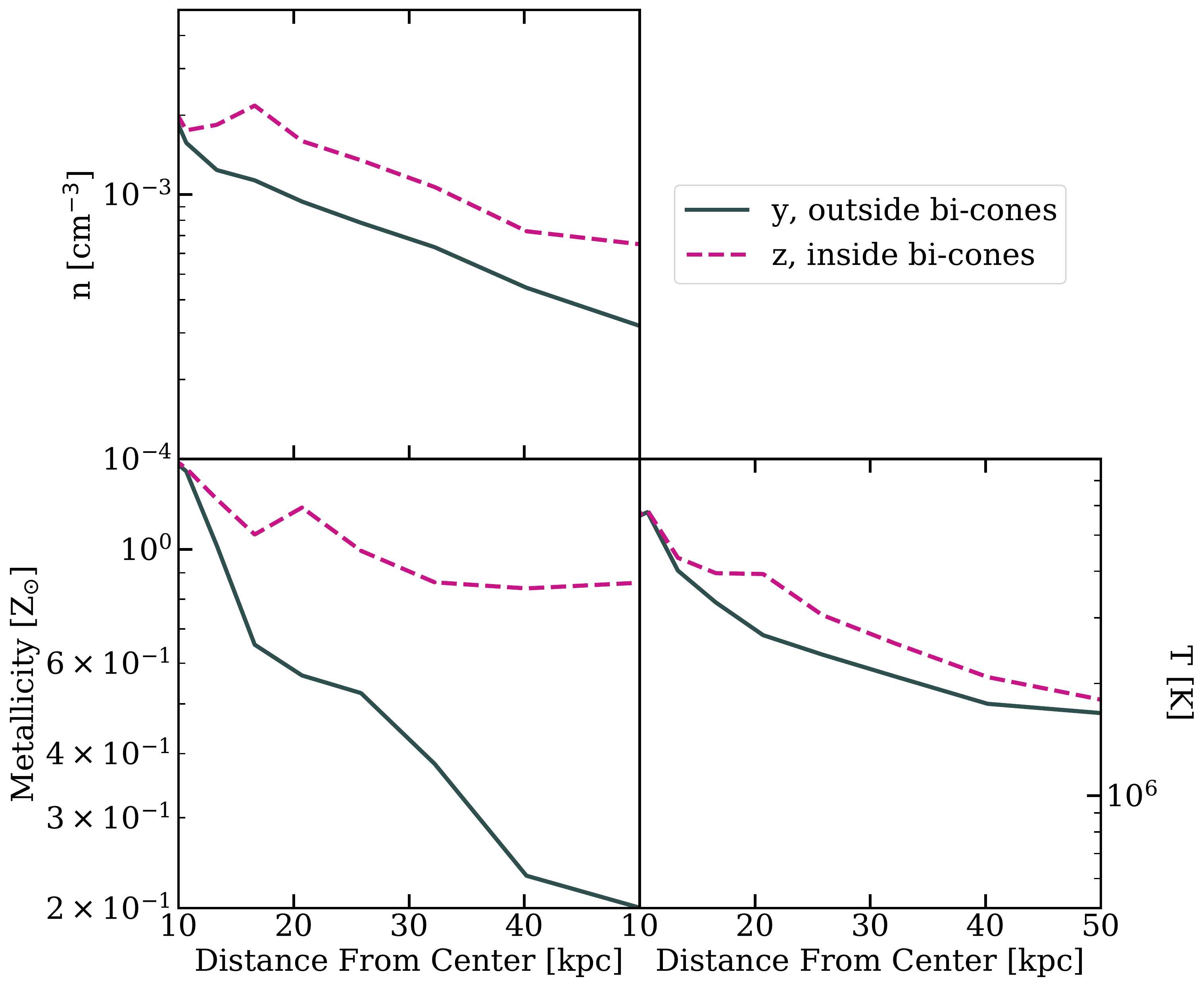}
    \caption{ Radial profiles of density, metallicity and temperature along the '$y$' (outside outflow cones) and '$z$' (inside outflow cones) directions for \hsfr. The quantities are weighted by X-ray luminosity ($0.5-2.0$ keV). 
    }
    \label{fig:dist_yz}
\end{figure} 

\begin{figure}

    \includegraphics[width=\columnwidth]{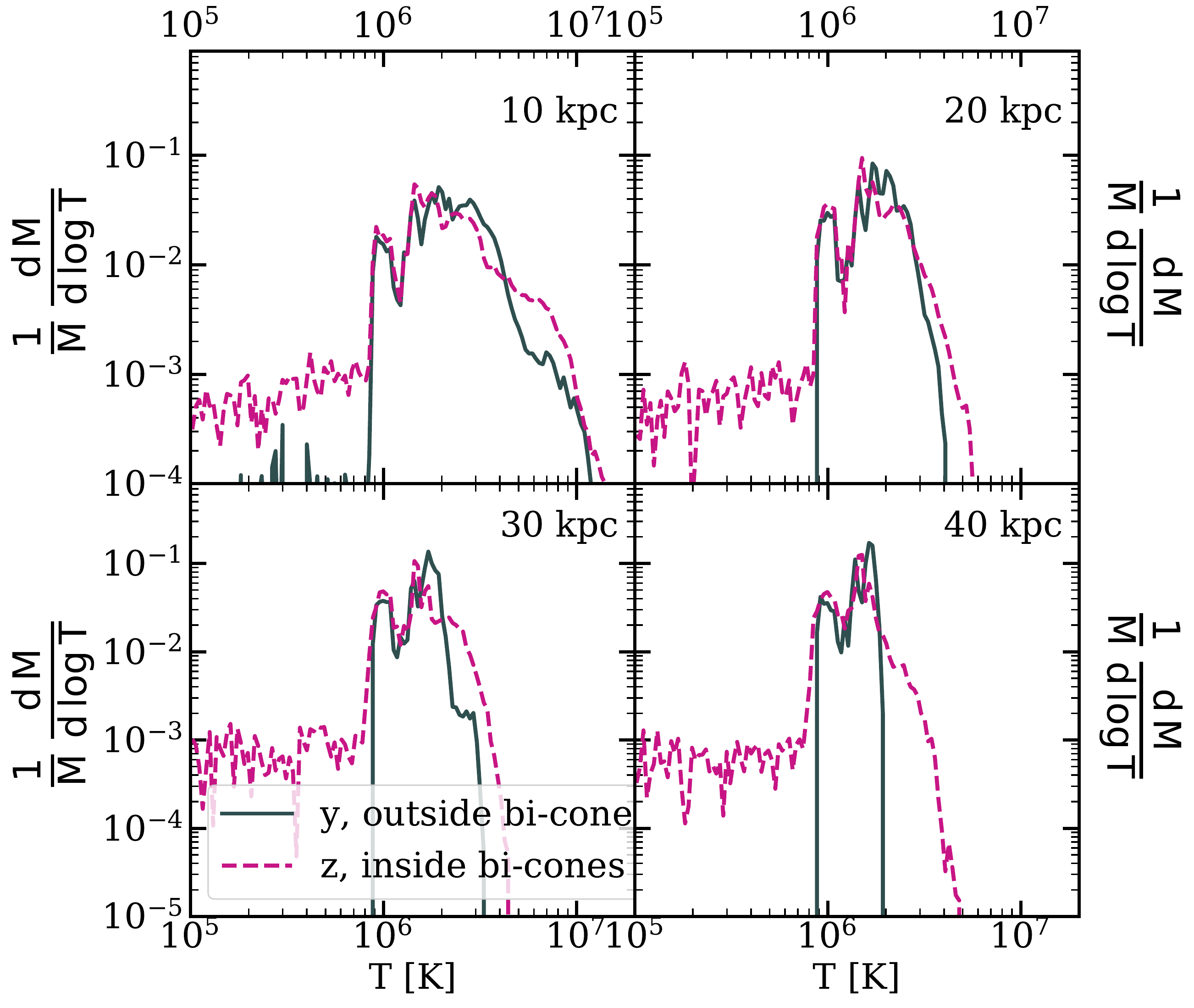}
    \caption{Comparison of temperature distribution in projected boxes ($800\times10\times 10$ kpc$^3$- the longest edge being along the $x$ axis). The panels show boxes at different distances from the centre of the galaxy, indicated at top-right corners. Inside the bi-cones, there is more gas with higher temperature, especially at large distance.
    }
    \label{fig:cellm_yz}
\end{figure}

From Figure \ref{fig:x-slices} we see that \hsfr\ has bi-conical outflows resulting different density, temperature and metallicity profiles inside and outside the bi-cones. Interestingly, these features are not obvious in the projected X-ray emission maps (see Figure \ref{fig:projected-emissivity}). Here, we investigate whether these outflow features produce visible differences in spectra taken from different azimuthal angles. High angular resolutions of future X-ray missions ($\sim 5''$ in Athena, for example) will allow astronomers to probe CGM at different azimuthal angles of a galaxy and provide information about profiles along and perpendicular to the outflow direction. Keeping this in view, in this Section we discuss the gas properties and spectra inside and outside the bi-cones, i.e., the `$z$' and `$y$' axes, respectively.

In Figure \ref{fig:dist_yz}, we show the variation of density, metallicity and temperature along `$y$' and `$z$' axes as the function of distance from the centre of the galaxy for \hsfr. To produce the profiles, we take rectangular boxes along $y$ and $z$ directions with dimensions $[x,y,z] \in[\pm10,\pm400,\pm10]$ kpc and $[x,y,z]\in[\pm10,\pm10,\pm400]$ kpc, centred at the origin of the simulation box. These boxes are denoted as ``y'' and ``z'' in the figure. We then average the quantities in the boxes along y- and z-direction, respectively. We note here that all the physical quantities shown here are weighted by $0.5-2.0$ keV X-ray luminosity. As expected from Figure \ref{fig:x-slices}, the average density, metallicity and temperature decrease away from the centre and are different in the two directions. Along the outflow direction we expect the gas temperature, density and metallicity values to be higher. The starkest difference is seen in the metallicity profiles along the two direction. For ``y'' direction, the metallicity is relatively flat around $Z_\odot$ for the inner $40$ kpc, whereas for the ``z'' direction, it drops fast, to around $0.2$ $Z_\odot$ (metallicity of pre-existing gas) at $40$ kpc. 

Figure \ref{fig:cellm_yz} shows the fractional mass distribution in temperature along the $z$- and $y$-directions for $10^{5-7.3}$ K gas, in a projected way. To produce these plots, we take rectangular boxes with dimensions $10\times10\times800$ kpc such that the longest dimension lies along the $x$ axis. We centre the box at different distances from the origin starting at $10$ kpc. For example, the centre of the box labelled `$10$ kpc' along $y$ direction is located at $[x,y,z]=[0,10,0]$ kpc. Note that this is different from gas sampling for Figure \ref{fig:dist_yz}.

The distributions along $y$-direction peak at $\sim 2 \times10^6$ K representing the less undisturbed medium which has not been affected by the outflows (other than the shocks). These distributions also become narrower for larger distances from the centre. In the $z$-direction, the high-temperature tail of the mass distribution is prominent,  especially at $30$ and $40$ kpc. The asymmetry of gas distribution in the two directions, generated by the bi-polar outflows, is clearly evident in this Figure.

\begin{figure}
    \includegraphics[width=\columnwidth]{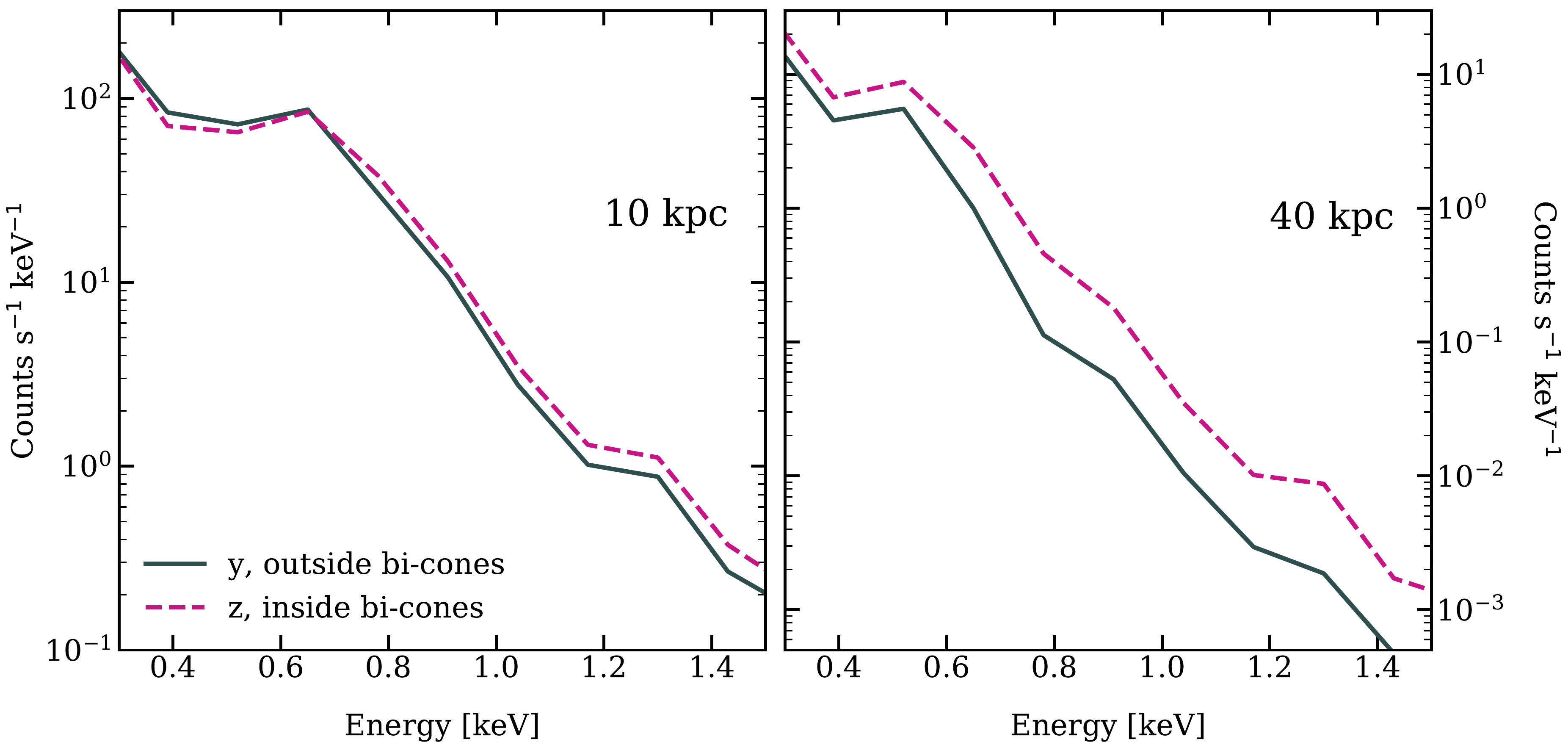}
    \caption{Spectra for boxes along $y$ and $z$ directions, binned at $130$ eV, located at a distance of $10$ and $40$ kpc, respectively. The spectra are nearly identical at $10$ kpc, while they differ significantly at $40$ kpc. The high-energy end is enhanced in the $y$ direction, due to larger amount of high-temperature gas.
    }
    \label{fig:spec_yz}
\end{figure}

In Figure \ref{fig:spec_yz} we show the spectra arising from boxes at $10$ and $40$ kpc. For producing the spectra, we use the projection of the boxes used for Figure \ref{fig:cellm_yz} along the $x$ axis, with arbitrary normalisations. We show only the low-resolution spectra binned at $130$ eV. At $10$ kpc the spectra from the two directions are nearly identical. This is not surprising as from Figure \ref{fig:cellm_yz}, the temperature distribution for X-ray emitting gas is nearly same at this distance from the centre. The difference in spectra is more prominent at $40$ kpc, especially at the high energy end. The spectra are less steep for the $z$-direction. This is because the $z$-direction has hotter gas at $40$ kpc than $y$-direction, as seen from Figure \ref{fig:cellm_yz}.

\begin{figure*}
    \includegraphics[width=\textwidth]{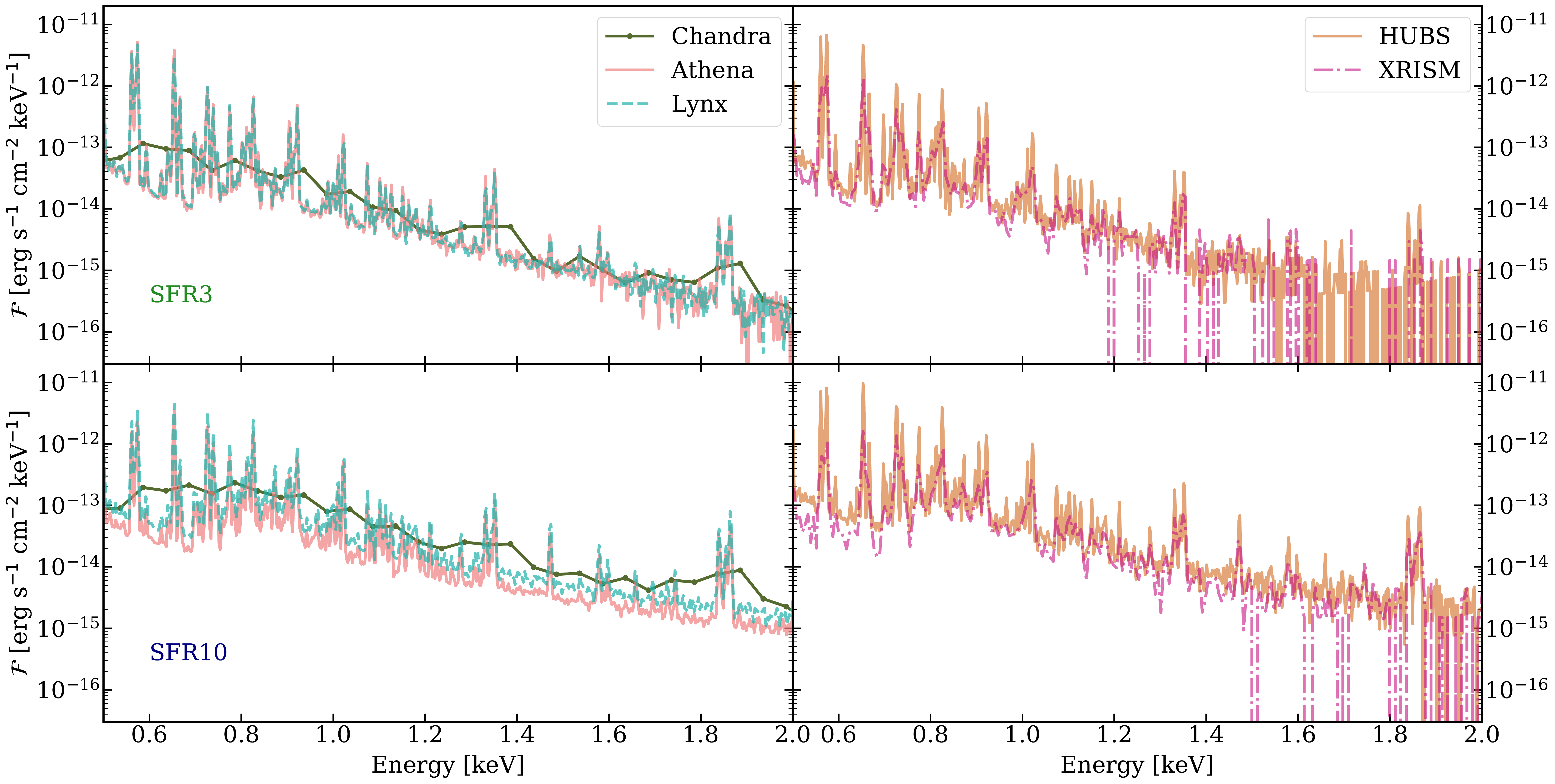}
    \caption{Mock spectra for Chandra (green), Athena (coral), Lynx (green), HUBS (yellow) and XRISM (magneta) spectra for \lsfr\ (top) and \hsfr\ (bottom). These spectra do not include background or foreground. The numbers on the top right corner indicate the $0.5-2.0$ keV luminosity in units of erg s$^{-1}$ calculated from these spectra. The spectral resolution of these instruments are $50.0$ (Chandra), $2.5$ (Athena), $3.0$ (Lynx) and $2.0$ (HUBS) eV. }
    \label{fig:mock_spec}
\end{figure*}

\section{Discussion}\label{sec:discussion}

\subsection{Mock Observations for Chandra, Athena, Lynx, HUBS \& XRISM}\label{subsec:predictions}
\begin{figure*}
	\includegraphics[width=\textwidth]{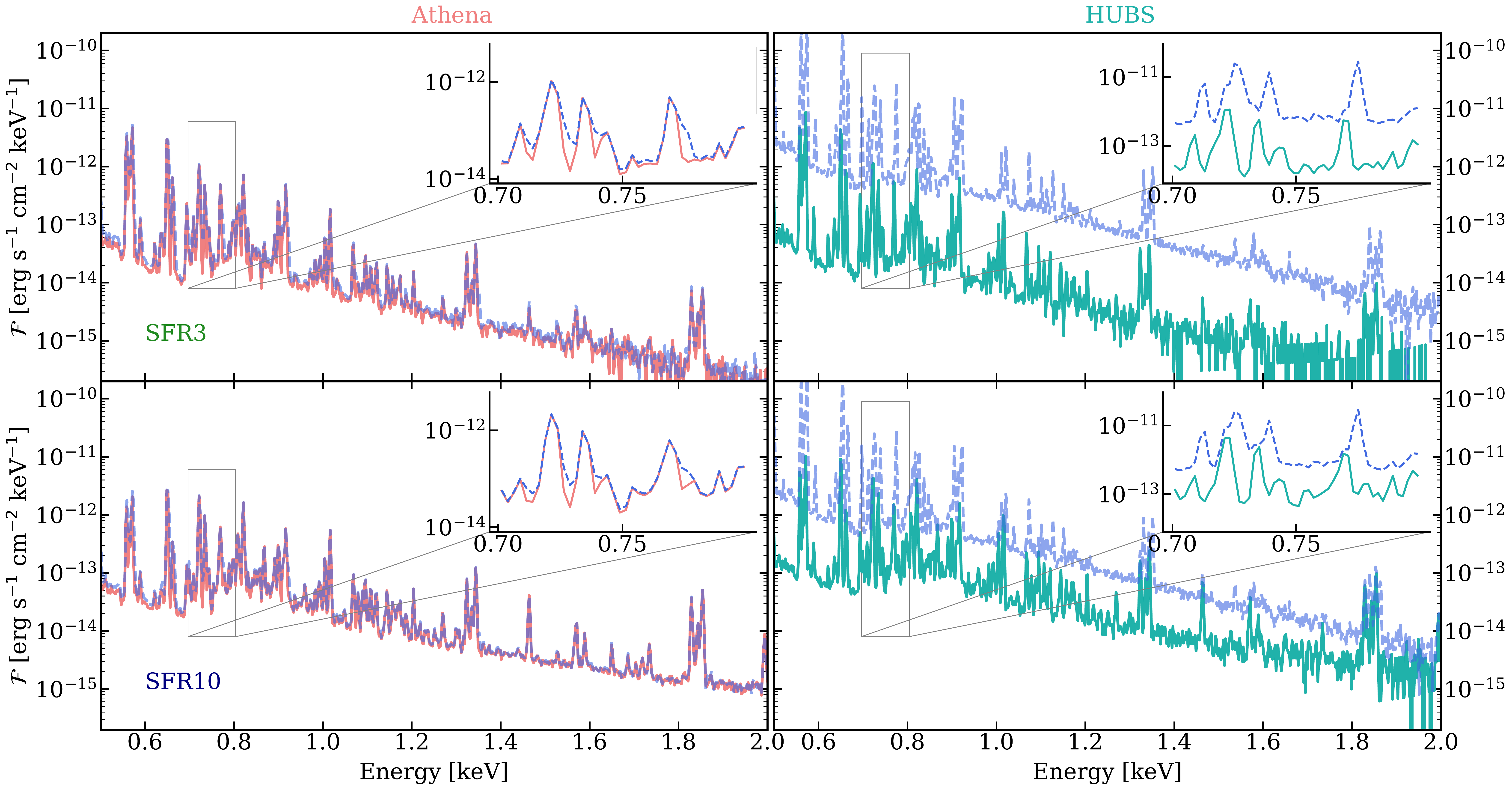}
	
    \caption{Mock spectra from Athena (left) and HUBS (right) for \lsfr\ and \hsfr\ at a distance of $25$ Mpc. We show the effect of adding Galactic foreground to the spectra in dashed lines. The spectra from HUBS exhibit a higher contribution from the foreground because of its large field of view ($\sim 1^{\circ}$), as compared to $5.9'$ field of view of Athena. In the top-right corner of each panel, we show the spectra zoomed-in between $0.7$ and $0.8$ keV. Emission lines at a distance of $25$ Mpc are not blended into the foreground for these eV-level instruments.
    } 
    \label{fig:mock_spectra_foreground}
\end{figure*}

In this Section, we discuss mock spectra produced from the simulation data. For our discussion, we focus on the following four instruments: ACIS-S on Chandra \citep{Chandra}, X-Ray Integral Field Unit on Athena \citep{Athena}, X-ray Microcalorimeter (Main Array) on Lynx \citep{Lynx}, the Hot Universe Baryon Surveyor (HUBS) \citep{Cui+20} and the X-ray Imaging and Spectroscopy Mission (XRISM) \citep{XRISM}. Except for Chandra, the rest of the telescopes are planned mission with enhanced capabilities \citep[see for a comparison]{Li2020}. These planned mission have a much higher energy resolution and sensitivity which will enable much more accurate measurements of the metallicity, gas kinematics, etc. 

We use a python-based module, SOXS\footnote{ \hyperlink{http://hea-www.cfa.harvard.edu/soxs/}{http://hea-www.cfa.harvard.edu/soxs/}.}, to create a photon list from the simulation data set assuming a distance of $25$ Mpc and an effective area of $2.5$ m$^2$. We choose this distance to ensure that the inner region of the CGM ($\lesssim 20$ kpc), which emits most strongly (see Figure \ref{fig:projected-emissivity}), fits within the field of view ($\sim 5'$) of Athena and Lynx\footnote{Chandra and HUBS have larger fields of view, $20'$ and $1^{\circ}$, respectively, while XRISM has a comparable field of view, $2.9'$.}. A subset of photons from the list are convolved with the instrument response in order to produce the spectra for an exposure time of $2$ Ms for Chandra and Athena and Lynx, and $7$ and $5$ Ms for HUBS and XRISM, respectively. We choose the exposure time for an instrument to minimise noise from the resulting spectra while ensuring that the photon list generated is not over-sampled.
We take the RA and Dec as $142.8^{\circ}$ and $+84.2^{\circ}$, respectively. The choice of the RA and Dec is arbitrary but it so happens that these coordinates correspond to those of NGC $4631$. We stress that our intention is not to compare the simulation spectra with this particular galaxy. . For \textit{Chandra}, the response files are for Cycle $22$. For Galactic absorption, we set the column density as $1.93\times 10^{20}$ cm$^{-2}$.

In Figure \ref{fig:mock_spec}, we show the flux obtained from each telescope for \lsfr\ (top) and \hsfr\ (bottom) between $0.5-2.0$ keV. Galactic foreground is not included. We note that Athena, Lynx, HUBS and XRISM, having a spectral resolution of $2.5$, $3.0$, $2.0$ and $5$ eV, respectively, are able to resolve line features currently unachievable by Chandra. All the three planned instruments are able to produce consistent spectra. The spectra generated using HUBS and XRISM are noisier than those from the other instruments.
The shape of the spectra from the three high-resolution instruments are comparable. We note here that because of their smaller field of view, the focus of Athena and Lynx will be on the smaller regions of galactic winds. XRISM will be able to resolve individual spectral lines, enabling observers to probe the multiphase structure of CGM gas \citep{XRISM}.

Figure \ref{fig:mock_spectra_foreground} shows the spectra from \lsfr\ (top) and \hsfr\ (bottom) for Athena and HUBS, adding Galactic foreground. The galaxy is placed at a distance of $25$ Mpc. We use SOXS to add the Galactic foreground, which is modeled as a uniform and diffuse plasma taken to be the sum of two thermal models at kT$=0.2$ and $0.099$ keV. The total foreground flux of an instrument depends on its field of view (FOV). As a result, the foreground flux of HUBS (FOV $\sim 1^{\circ}$) is much larger than that added to Athena (FOV $\sim 5'$). For HUBS, in the softer energy band, the foreground is $\sim 15$ times larger than the signal itself. In the top-right corner of each panel in Figure \ref{fig:mock_spectra_foreground}, we also show the zoom-in spectra between $0.7-0.8$ keV. There we can see that the lines from the external CGM are redshifted relative to the foreground. This indicates that at the eV-level spectral resolution, the emission lines from the CGM at this relatively close distance ($25$ Mpc) are not blended into the foreground.

 \subsection{Connection to Other Works}

The 1-T/2-T model have been used to fit the spectra of many nearby star-forming galaxies \citep{Dahlem+98, Tullmann+06, Yamasaki+09, Owen+09, Mineo+12, Bogdan+13,Lopez+20}. Our best-fit temperatures are broadly consistent with the reported values. For MW, a hot component has been detected close to the virial temperature of the halo, $\sim (2-3)\times$\mk, in emission \citep{Henley+13, Nakashima+18} as well as absorption \cite{Hagihara+10, Gupta+12,fang15, Gupta+17, Das+19b}. We find that the higher fitting temperatures, T$_{\rm 2T, high}$, which are $2.8$ and $3 \times$\mk respectively for \lsfr\ and \hsfr, are consistent with the observations. That said, as we emphasize in previous sections, the best-fit temperatures using 1-T/2-T models may not reflect faithfully the underlying temperature distribution. Theoretically, it may be possible to find several mathematical functions, including 3-T model \citep{Lopez+20}, that fit the spectra even better than the proposed log-normal. However, not all of these functions will be able faithfully reflect the temperature distribution of the hot CGM, which from the simulations we know is close to a log-normal.

\begin{figure}
	\includegraphics[width=\columnwidth]{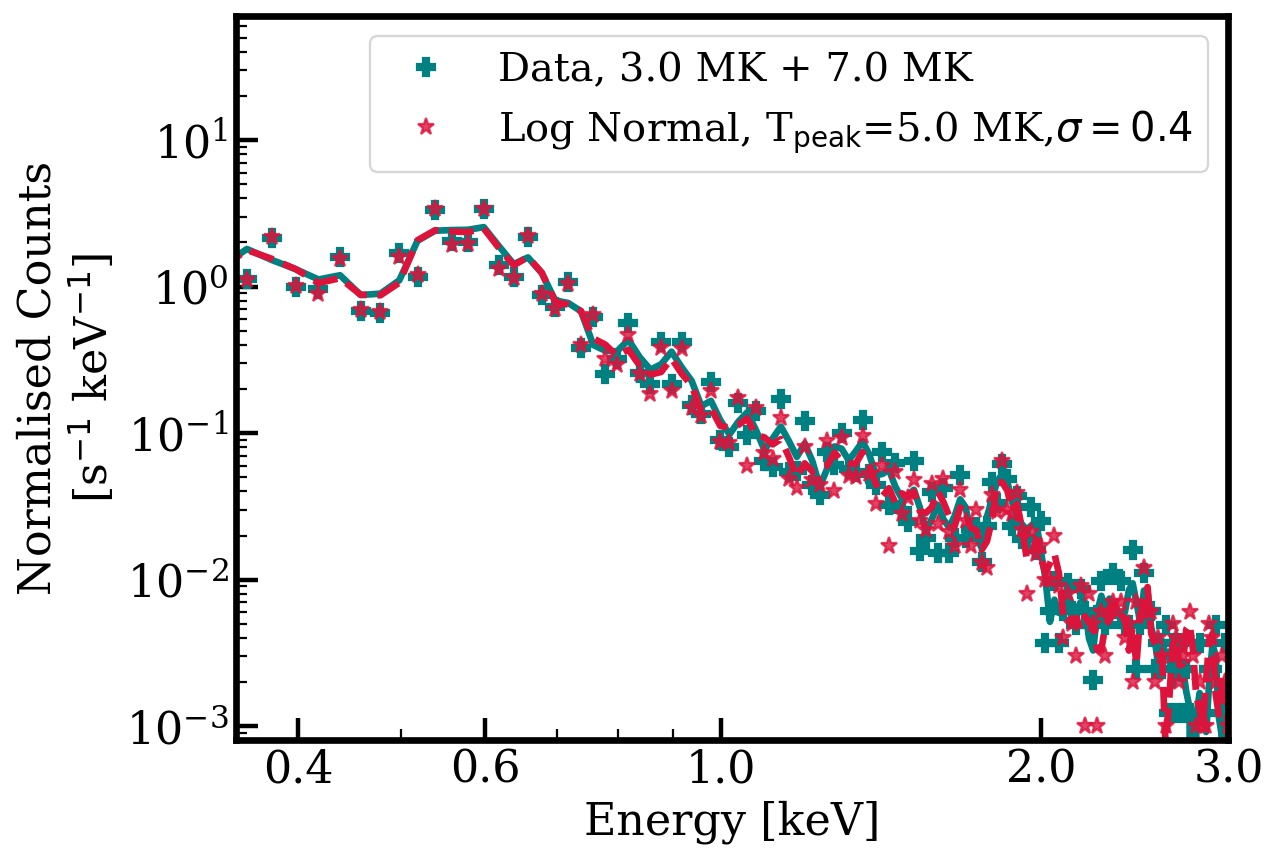}
    \caption{ The spectrum for NGC $4631$ using fitting parameters given in \citet{Li+13} compared with a log-normal box. We reproduce the \textit{Chandra} spectrum for NGC $4631$ from \citet{Li+13} (teal) by 
    superposing normalised spectra from two uniform temperature boxes at $3 \times 10^6$ and $7\times10^6$ K, which are obtained by 2T model fitting reported by \citet{Li+13}. In green we show the best-fit spectrum from the log-normal model with a peak at $5 \times 10^6$ K. The symbols represent the mock data generated using SOXS, while the lines represent a fit to the data points. 
    } 
    \label{fig:logn_chandra}
\end{figure}

To illustrate the applicability of the log-normal model to existing data, we discuss the \textit{Chandra} spectrum of NGC $4631$. \cite{Wang+01} and \cite{Li+13} discuss the \textit{Chandra} spectrum for NGC $4631$ and fit it using the 2T model at $\sim 3\times 10^6$ and $\sim 7\times 10^6$ K, having luminosities equal to $9.3\times 10^{38}$ and $13.7\times 10^{38}$ erg s$^{-1}$, respectively. Figure 3 of \cite{Wang+01} shows that the 2T model can fit the data quite well. 

We reproduce the spectrum using these temperatures in Figure \ref{fig:logn_chandra} (teal solid line) following the procedure detailed in Section \ref{subsec:1T-2T-spec}. 
We convolve this spectrum with \textit{Chandra} instrument response corresponding to Cycle $0$ to generate a $60$ ks observation, same as that used for the actual observation in \cite{Li+13}. We take the values for RA and Dec, the column density, and the distance estimates from Table $1$ of \cite{Li+13}.

In order to compare with the log-normal model, construct log-normal temperature box by assuming a peak temperature and a width of the log-normal distribution curve. We produce the \textit{Chandra} mock spectrum from such a distribution for $60$ ks observation period and the relevant parameters for NGC $4631$. 
We find that the best-fit peak temperature is $5\times 10^6$ K with a corresponding $\sigma=0.4$, and we show the best-fit log-normal spectrum in Figure \ref{fig:logn_chandra} (red dashed line). We note that log-normal spectrum matches the observed spectrum well. This shows that log-normal model is able to fit the existing observations as well. 

Observations also indicate that models beyond a simple 1-T/2-T is required. For example, broad distributions of temperature have been used to explain the emission \citep{Strickland+02} and absorption studies \citep{Gissis+20} where multiple ions and ionization states are present. 
Spectra from galaxy clusters also has been interpreted using multi-temperature fits \citep{Sanders+10, Werner+13, Hitomi-collab}. The log-normal temperature model has also been used in other context for fitting X-ray spectra. For example, \citep{Ge+15} model the X-ray emission from unresolved stellar contributions using a log-normal distribution. Gaussian temperature distributions have been used previously for fitting X-ray spectra from clusters \citep{Kaastra+04, Simionescu+09}. In the context of groups and clusters, we do expect that similar physical processes, which determine the log-normal shape of the temperature distribution of the gas in our simulations, will be operating and can leave the underlying temperature and density distributions broadened. The high spectral resolution (about $49$ eV at $0.3$ keV) of the eROSITA could potentially detect these distributions in its targets \citep{eRosita}.

Log-normal distribution has been discussed in recent theoretical studies of CGM as well. Several simulations, evolving the CGM both in the cosmological context and in an isolated environment, found gas temperature and density have a broad distribution at different radii \citep{Lochhaas+20, Lochhaas+21, Fielding+SMAUG}. Physically, a broadened distribution in temperature can arise because of several factors, such as the interaction of enriched gas with the ambient medium, adiabatic expansion of hot outflows, the interaction of outflows from multiple star formation events, radiative cooling, turbulent motion, etc \citep{Blaisdell+93}.  Further, gas models obeying log-normal distributions are also used to explain the observations of emission and column densities of oxygen lines \citep{Faerman+17, McQuinn&Werk18, Voit19,faerman20}. 

\section{Summary}\label{sec:summary}

Hot galactic outflows emerge as a result of star formation. They interact with the ambient medium and produce multi-phase gas in the CGM of the galaxy. The outflows from \lsfr\ form a large-scale fountain and nearly isotropic CGM, while those in \hsfr\ expand bi-conically along the vertical direction.

The main conclusions of the Paper are the following:

\begin{enumerate}[leftmargin=\parindent]
    \item The X-ray emitting CGM exhibits a broad range of temperatures, density and metallicity (Figure \ref{fig:x-slices}, \ref{fig:denT-histo}).  

    \item The broad temperature range of the X-ray emitting CGM cannot be characterised using 1-T/2-T models, even though such models may fit the X-ray emission spectra (Figure \ref{fig:spectra-all}, \ref{fig:1T-mass-temperature}). 
    
    \item The mass distribution of X-ray emitting CGM can be better approximated by a log-normal distribution in temperature for both low and high SFR. Such a distribution can be described using a two parameters, i.e, a peak temperature and a width around the peak (Eq. \ref{eqn:log-normal}).
   
    \item The log-normal distribution fits better the spectra than 1-T/2-T model, and best-fit parameters reproduce the underlying temperature distribution (Figure \ref{fig:spectra-all}, \ref{fig:chi}, \ref{fig:ln-td}).
    
    \item In the high SFR case, \hsfr, the outflows introduce an asymmetry in gas properties inside and outside of the outflow bi-cones. This difference is reflected in the X-ray spectra, as gas inside the cone is hotter and more metal-enriched (Figure \ref{fig:dist_yz}, \ref{fig:cellm_yz}, \ref{fig:spec_yz}). 
    
    \item Future X-ray telescopes with $\sim$eV resolution, such as Athena, Lynx and HUBS, will be able resolve the line features, providing much more details about the hot CGM (Figure \ref{fig:mock_spec},\ref{fig:mock_spectra_foreground}).

\end{enumerate}

\section*{Acknowledgements}
We dedicate this paper to all the frontline workers across the world, who have worked so hard to keep us safe during the challenging time of COVID-19. We thank the anonymous referee for their useful comments. ML thanks the beneficial discussion with S. Mathur, J-T Li, D. Wang, S. Das, and L. Lopez. We thank the KITP-Halo21 program (supported by NSF grant PHY-1748958) for stimulating scientific discussions. Data analysis and visualization are partly done using the \textsf{yt} project \citep{turk11}. Part of the computation/analysis are performed on the Rusty cluster of the Simons Foundation. We thank the Scientific Computing Core of the Simons Foundation for their technical support. A part of the analysis was also done at SahasraT cluster, maintained by the Super-computer Education and Research Centre, Indian Institute of Science, and the cluster at Shanghai Astronomical Observatory. AV would like to thank the staff at both these institutes for their support. 
\section*{Data availability}
The data underlying this article will be shared on reasonable request to the corresponding author.



\bibliographystyle{mnras}


\appendix

\section{Temporal Variations in the Mass Distribution}\label{app:temporal_variation_mass_dist}

\begin{figure}
	\includegraphics[width=\columnwidth]{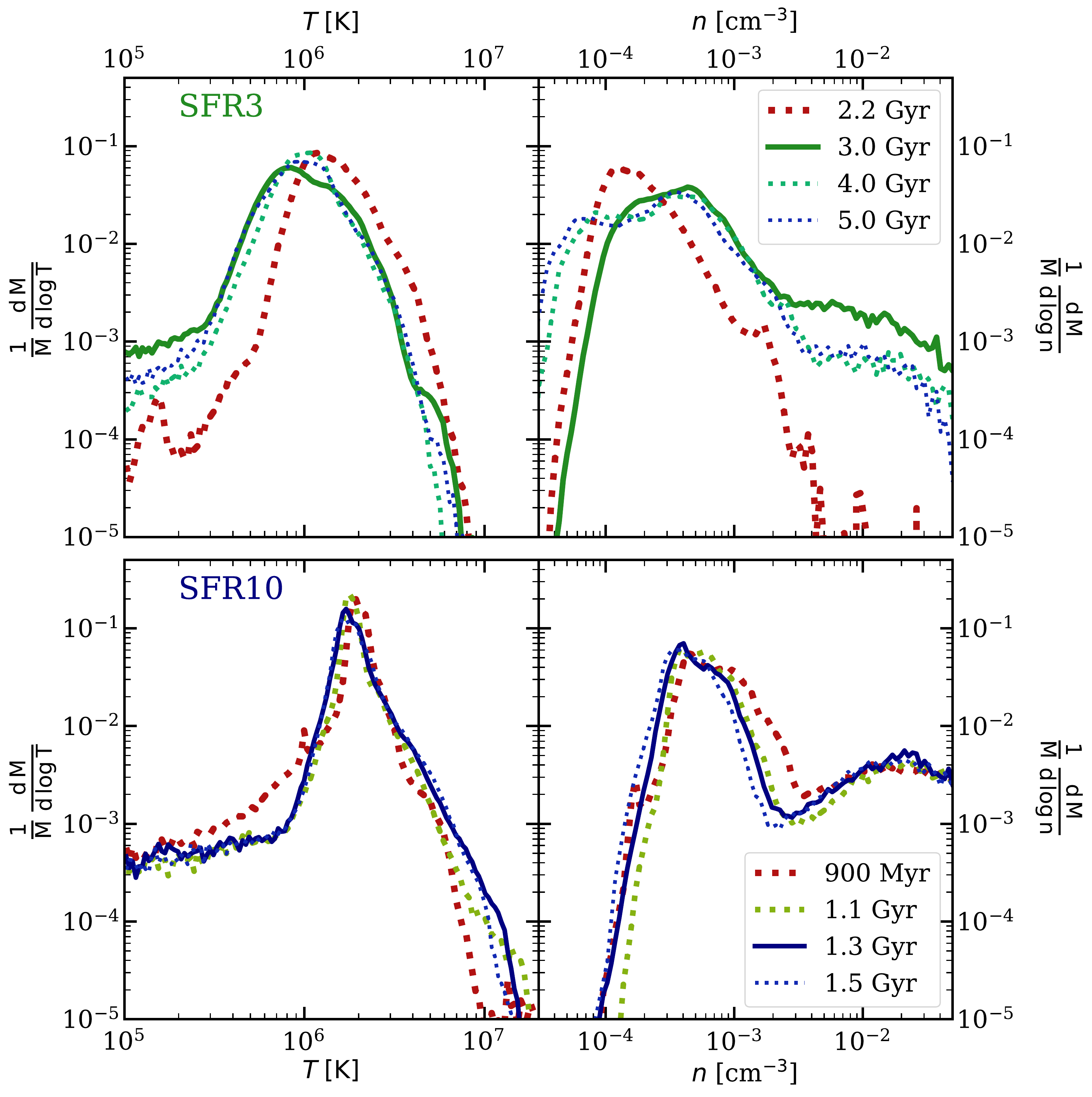}
	
    \caption{Mass probability distribution for logarithmic density and temperature for \lsfr\ (top) and \hsfr\ (bottom). Each curve represents a different time step in the simulations. The systems reach a quasi steady state in terms of mass distribution. Maximum mass is at $\sim 10^6$ K for all the times. The solid green and blue curves represent the time steps analysed in the main body of the paper.
    }
    \label{fig:t-series}
\end{figure}

In previous Sections, we discuss the results from one snapshot of the simulation runs. However, the underlying density and temperature distributions of the gas does not vary significantly over time after the system reaches a steady state.

In Figure \ref{fig:t-series}, we show 1-D mass plots for different time steps for \lsfr\ and \hsfr. The solid curves, at $3$ Gyr for \lsfr\ and $1.3$ Gyr for \hsfr, are identical to the ones analysed in the main body of the paper. We note that in the \lsfr\ case, condensation of cool gas begins in the CGM at about $2.2-3$ Gyr and the system settles into a quasi-steady state thereafter. For the \hsfr case the hot CGM attains a quasi-steady state (for a few $100$s Myr) after $700$ Myr. The variations in the density and temperature profiles that we see in Figure \ref{fig:t-series} produce only a factor of $2$ variation in the spectra, as shown in Figure \ref{fig:spectra-all}.

\section{Variations in Sizes of selected CGM}\label{sec:box-size}

Our analysis is limited to the inner $R\lesssim 50$ kpc of the simulation domain, which we refer to as the ``region of interest''. We check how varying the size of the included region change our results, including both the inner and outer boundaries. 

\begin{figure}
	\includegraphics[width=\columnwidth]{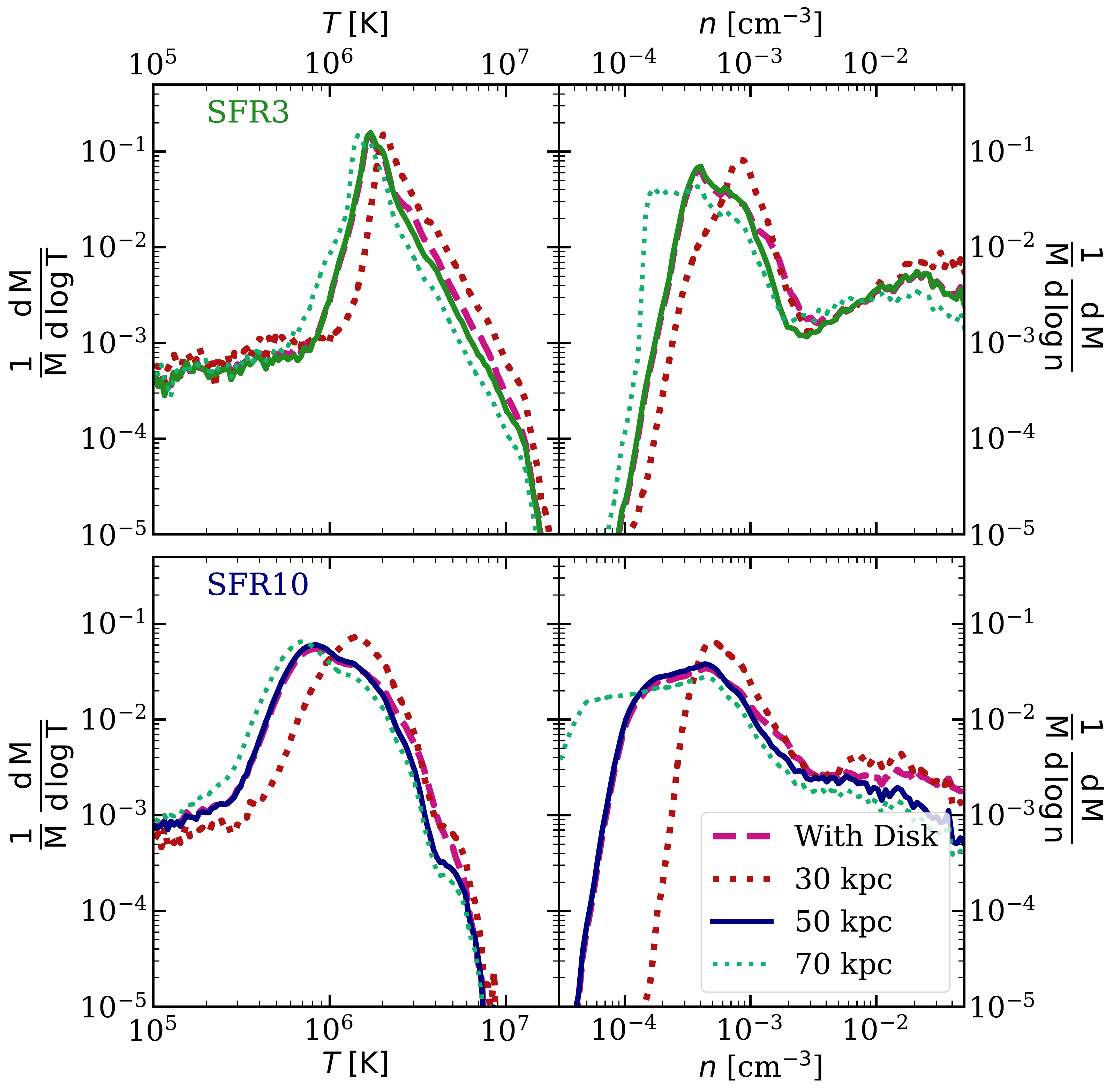}
	
    \caption{Mass probability distribution for logarithmic density and temperature for \lsfr\ (top) and \hsfr\ (bottom). Each curve represents a different size of selected CGM whose dimension is indicated at the bottom right corner of the figure. The curve corresponding to ``With Disk'' refers to a $50$ kpc region which included the disk. Mass distributions do not vary much when changing the size of the selected region. 
    }
    \label{fig:box-var-masst}
\end{figure}

Figure \ref{fig:box-var-masst} shows the fractional mass distribution of temperature (left) and density (right) for \lsfr\ (top) and \hsfr\ (bottom). This figure is identical to Figure \ref{fig:t-series}, except that now we show the fractional mass distribution when we consider regions of different sizes. We show the outer radii of $30$, $50$ (fiducial value) and $70$ kpc, respectively. In all these cases we have removed the disk region ($\pi 20\times20\times (\pm 3$) kpc$^3$) from the analysis using the procedure described in Section \ref{subsec:rho-T-histos}. The distributions are nearly identical. We also show the fractional mass distribution for a region that includes the disk region in each panel, i.e. all gas within the outer boundary (dashed curve). The fractional mass distribution does not change significantly. The differences in fractional mass distribution of density of different sizes are noticeable, especially between $30$ and $70$ kpc. This difference arises because large regions include relatively higher proportion of low density gas. The regions with low density regions do contribute significantly to the X-ray emission because of smaller density and because they lie at large radii, far away from the region of interest.

\begin{figure}
	\includegraphics[width=\columnwidth]{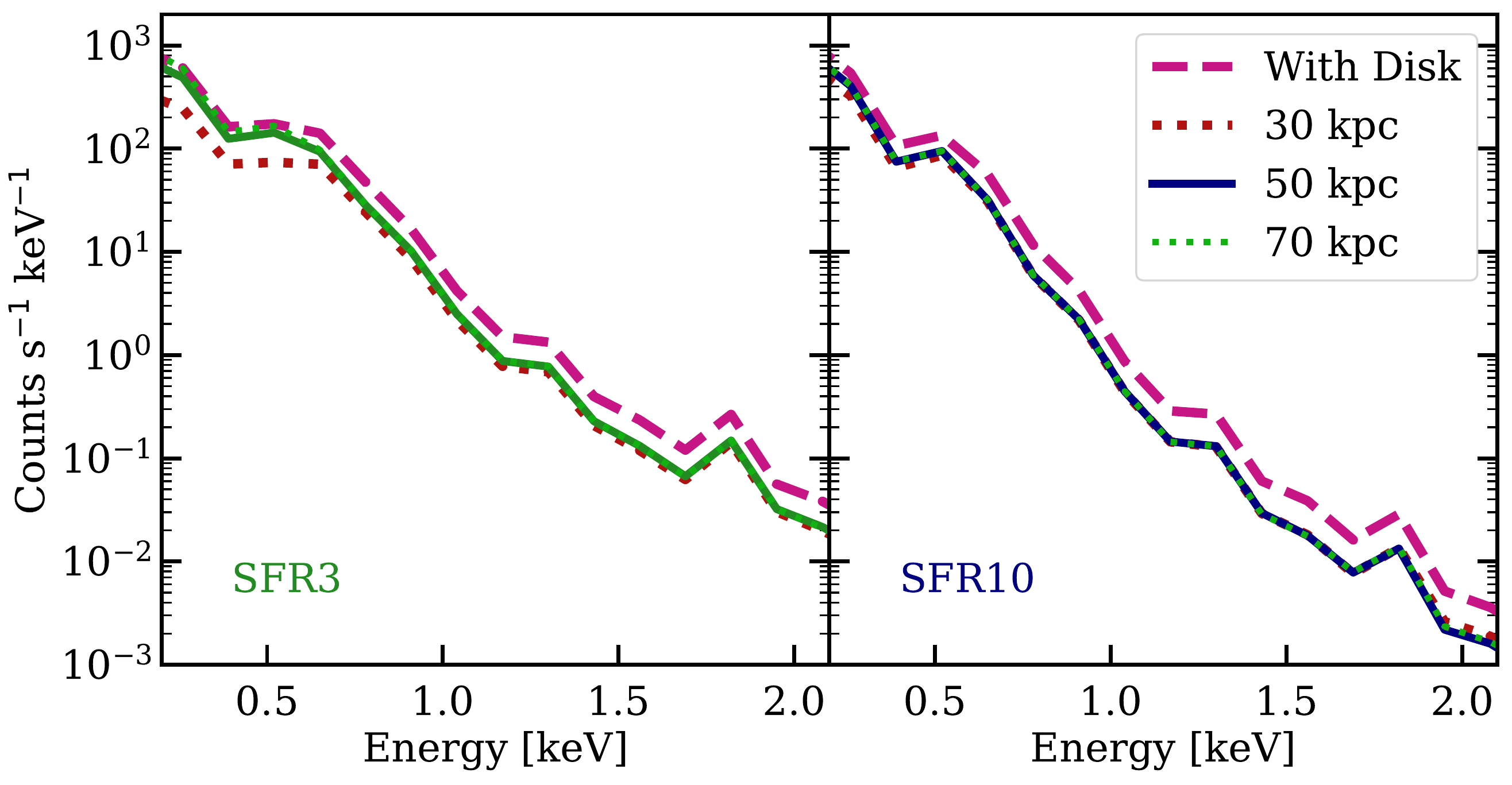}
	
    \caption{Low-resolution spectra for \lsfr\ (left) and \hsfr\ (right). Each curve represents a different box size whose dimension is indicated at the bottom right corner of the figure. The solid curves of the same color represent those in Figure \ref{fig:denT-histo}. The variation of the spectra with different sizes of selected regions is minor.
    } 
    \label{fig:box-var-specta}
\end{figure}

Figure \ref{fig:box-var-specta} shows the low-resolution spectra corresponding to each of the different box sizes, which do not show significant variation, either. This is not surprising since the underlying mass distribution does not vary significantly by changing the box size. Thus, we conclude that our results are not sensitive to the selected region of interest.

\label{lastpage}
\end{document}